\documentclass[twocolumn]{aastex63}

\usepackage{amsmath}
\usepackage{amssymb}
\usepackage{rotating}

\newcommand\Fs{F_{\rm s}}
\newcommand\Fb{F_{\rm b}}

\newcommand\tE{t_{\rm E}}

\newcommand\piEN{\pi_{\textrm{E},N}}
\newcommand\piEE{\pi_{\textrm{E},E}}

\newcommand\muhelio{\boldsymbol{\mu}_{\rm rel}}

\newcommand\ve{\boldsymbol{v_{\oplus,\perp}}}

\newcommand\thetaE{\theta_{\rm E}}
\newcommand\piE{\pi_{\rm E}}
\newcommand\piEvec{\boldsymbol{\pi_{\rm E}}}

\newcommand\OB{OGLE-2011-BLG-0462}

\received{}
\revised{}
\accepted{}

\shorttitle{Systematic errors in OGLE-2011-BLG-0462}
\shortauthors{P. Mr\'oz et al.}

\begin{document}

\title{Systematic Errors as a Source of Mass Discrepancy in Black Hole Microlensing Event OGLE-2011-BLG-0462}

\correspondingauthor{Przemek Mr\'oz}
\email{pmroz@astrouw.edu.pl}

\author[0000-0001-7016-1692]{Przemek Mr\'oz}
\affil{Astronomical Observatory, University of Warsaw, Al. Ujazdowskie 4, 00-478 Warszawa, Poland}

\author[0000-0001-5207-5619]{Andrzej Udalski}
\affil{Astronomical Observatory, University of Warsaw, Al. Ujazdowskie 4, 00-478 Warszawa, Poland}

\author{Andrew Gould}
\affil{Max Planck Institute for Astronomy, K\"onigstuhl 17, D-69117 Heidelberg, Germany}
\affil{Department of Astronomy, The Ohio State University, 140 W. 18th Avenue, Columbus, OH 43210, USA}

\begin{abstract}
Two independent groups reported the discovery of an isolated dark stellar remnant in the microlensing event OGLE-2011-BLG-0462 based on photometric ground-based observations coupled with astrometric measurements taken with the \textit{Hubble Space Telescope}. These two analyses yielded discrepant mass measurements, with the first group reporting that the lensing object is a black hole of $7.1 \pm 1.3\,M_{\odot}$ whereas the other concluded that the microlensing event was caused by either a neutron star or a low-mass black hole ($1.6\!-\!4.4\,M_{\odot}$). Here, we scrutinize the available photometric and astrometric data and conclude that systematic errors are a cause of the discrepant measurements. We find that the lens is an isolated black hole with a mass of $7.88 \pm 0.82\,M_{\odot}$ located at a distance of $1.49 \pm 0.12$\,kpc. We also study the impact of blending on the accuracy of astrometric microlensing measurements. We find that low-level blending by source companions is a major, previously unrecognized, challenge to astrometric microlensing measurements of black hole masses.
\end{abstract}


\section{Introduction} \label{sec:intro}

Using gravitational microlensing to detect ``unobservable bodies'' was discussed as early as in the 1960s by \citet{liebes1964}. However, despite the rapid progress of microlensing experiments in the past decades, it was not until recently that the first robust discovery of an isolated stellar remnant in the microlensing event \OB{} was announced by two independent groups \citep{sahu2022, lam2022, lam2022_supl}. The \textit{late} discovery of \OB{} is rooted in the difficulty of measuring the masses of dark lenses in single-lens microlensing events. The mass of the dark lens $M$ and its distance $D_l$ can be determined only if two quantities, the angular Einstein radius $\thetaE$ and the microlensing parallax $\piE$, are measured:
\begin{equation}
M = \frac{\thetaE}{\kappa\piE};\quad \frac{\mathrm{au}}{D_l}= \piE\thetaE+\frac{\mathrm{au}}{D_s}
\end{equation}
where $\kappa=8.14\,\mathrm{mas}\,M_{\odot}^{-1}$, and $D_s$ is the distance to the source star. These two quantities are very rarely measured together in single-lens microlensing events that have only photometric observations. 

Measurements of the angular Einstein radius are a key problem: they require almost a perfect (and rare) alignment between the lens and the source to detect finite-source effects in the light curve of the event \citep{gould1994,witt1994,nemiroff1994}. This situation can be ameliorated with astrometric or interferometric observations of microlensing events, which may allow one to directly measure the angular Einstein radius, even in the absence of finite-source effects. Precise astrometric measurements may enable one to detect subtle movement of the image centroid in the sky as a result of gravitational microlensing \citep{miyamoto1995,hog1995,walker1995}, whereas interferometric observations allow one to measure the separation between the two images of the source created during the lensing event \citep{delplancke2001}. The amplitude of these effects ($\lesssim 1\,\mathrm{mas}$) is at the hairy edge of the capabilities of current instruments.

Gravitational microlensing event \OB{} was detected on 2011 June 2 almost simultaneously by two surveys: OGLE and MOA (as MOA-2011-BLG-191). In 2011 August, when it was realized that the event is a promising black hole candidate, astrometric observations with \textit{Hubble Space Telescope} (\textit{HST}) were initiated by a group led by K.~Sahu. Although the microlensing astrometric deviation is expected to be largest near the peak of the event (in high-magnification events at $\approx t_0 \pm \tE \sqrt{2}$, where $t_0$ is the peak time, and $\tE$ is the Einstein radius crossing timescale), additional observations taken well after the event are needed to disentangle microlensing astrometric effects from the proper motion of the source star. Sahu's group obtained eight \textit{HST} observations that span 2011--2017. Meanwhile, when \textit{HST} images of \OB{} became public, the event caught the attention of another group led by C.~Lam and J.~Lu, who secured an additional \textit{HST} epoch in the late 2021. The two groups worked independently of each other and published their results separately.

The analyses of both groups were based almost on the same \textit{HST} images. \citet{lam2022} used one additional \textit{HST} epoch from 2021 October, but did not use one image from 2013 May, which was incorporated in the \citet{sahu2022} analysis. \citet{lam2022} used only OGLE photometric data, whereas \citet{sahu2022} used both OGLE and MOA data to constrain the model parameters. The \textit{HST} data were reduced using different methods. One of the challenges possibly impacting the astrometric measurements is the presence of a bright neighbor star located only 0.38\,arcsec to the west of the source star. \citet{lam2022} carried out injection and recovery tests to check for possible systematic shifts that may affect their astrometric measurements. They found a bias toward the neighbor in later epochs, when the target was no longer significantly magnified. The bias has a range of values in different orientations; see Table~16 of \citet{lam2022_supl}. The mean bias in the west direction is $0.38 \pm 0.07$ and $0.10 \pm 0.09$\,mas for the $F814W$ and $F606W$ filters, respectively. The bias in the north--south direction has a lower value and varies from $-0.185$ (toward the south) to $+0.298$\,mas (toward the north), depending on the epoch. \citet{sahu2022} developed a special software to subtract the \textit{HST} point spread function (PSF) from the neighboring star and to measure the position of the source. As we show in Sections~\ref{sec:seeing} and~\ref{sec:hst}, the measurements taken by \citet{lam2022} and \citet{sahu2022} are inconsistent with each other. As we have little expertise in the analysis of the \textit{HST} data, we are a priori agnostic as to which reductions are ``more correct.''

Both groups reported a successful detection of astrometric microlensing effects and the measurement of the angular Einstein radius of the lens of $5.18 \pm 0.51\,\mathrm{mas}$ \citep{sahu2022} and $3.89^{+1.12}_{-1.16}\,\mathrm{mas}$ \citep[][their default weight, hereafter DW, model]{lam2022}. Similarly, the reported masses of the lens are different: $7.1 \pm 1.3\,M_{\odot}$ \citep{sahu2022}, $3.79^{+0.62}_{-0.57}\,M_{\odot}$ \citep[][DW model]{lam2022}, and $2.15^{+0.67}_{-0.54}\,M_{\odot}$ \citep[][equal weight, hereafter EW, model]{lam2022}; these differences are due to both different angular Einstein radii and different microlensing parallaxes inferred by the two groups.

\begin{figure}
\includegraphics[width=.5\textwidth]{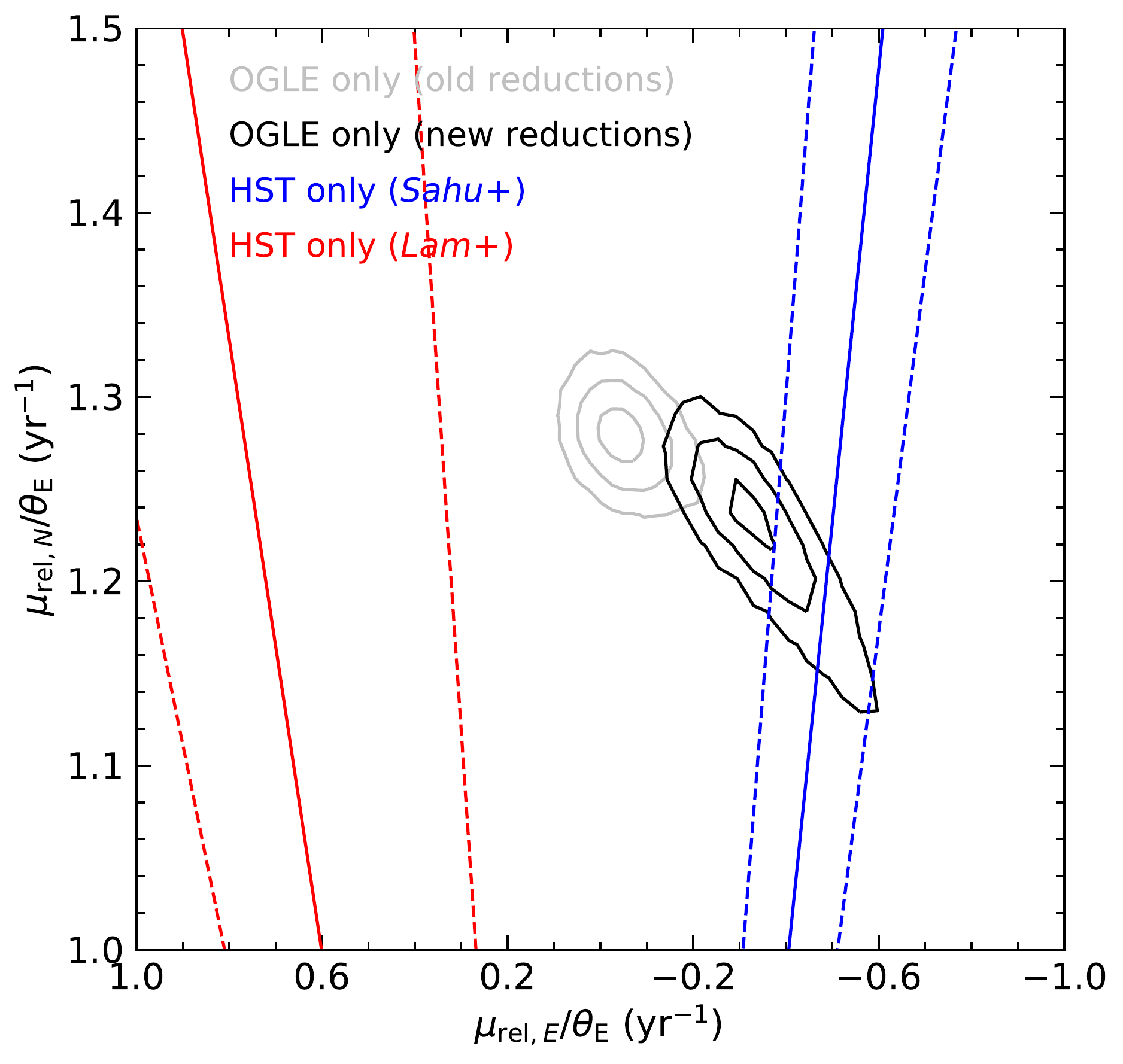}
\caption{Constraints on the direction of the heliocentric relative lens-source proper motion derived from OGLE photometric data (gray and black contours) and \textit{HST} astrometric data (blue and red lines). Contours represent 39.3\%, 86.5\%, and 98.9\% confidence regions. Solid and dashed lines correspond to the best-fit direction and its $1\sigma$ error bar, respectively.}
\label{fig:pm1}
\end{figure}

These differences are partly caused by a tension between photometric and astrometric measurements, which was noticed (and handled differently) by both groups. Both photometric and astrometric observations carry information about the direction of lens-source relative proper motion. For example, photometric data allow one to measure the microlensing parallax vector $\piEvec$, which is related to the heliocentric relative lens-source proper motion $\muhelio$ via the following:
\begin{equation}
\frac{\muhelio}{\thetaE} = \frac{1}{\tE} \frac{\piEvec}{\piE}+\frac{\ve\piE}{\mathrm{au}},
\end{equation}
where $\ve(N,E)=(-2.01,\ 25.41)$\,km\,s$^{-1}$ is the Earth's velocity in the north and east directions projected on the sky at the time of the closest lens-source approach $t_0$ \citep{gould2004}. The direction of $\muhelio$ can also be read off the astrometric data.

\citet{sahu2022} noticed that the direction of $\muhelio$, parameterized by an angle $\varphi$, is different when derived using photometry alone ($\varphi=353.7 \pm 2.3$\,deg) and astrometry alone ($\varphi=337.9\pm5.0$\,deg). This tension is illustrated in Figure~\ref{fig:pm1}, which shows constraints on the direction of $\muhelio$ derived using OGLE photometry (gray contours) and \textit{HST} astrometry \citep[blue and red lines;][]{sahu2022}. \citet{sahu2022} also noticed a correlation between $\varphi$ and the magnitude of the parallax vector $\piE$; they scaled the photometric errors ``by different amounts in different temporal bins making sure that the scaled errors are compatible with the statistical dispersions in the measurements,'' effectively increasing the size of the OGLE error ellipse. Then, in the $x$ (east--west) direction, the OGLE-ellipse error is larger than the \textit{HST} (1D) error, so the \textit{HST} data dominate. In the $y$ (north--south) direction, there is essentially no \textit{HST} information, so OGLE dominates (see Figure~\ref{fig:pm1}). Their final model has $\varphi=342.5 \pm 4.9\,\mathrm{deg}$ and $\piE = 0.0894 \pm 0.0135$, and so the small parallax, in conjunction with a relatively large $\thetaE$, drives the lens mass to $\sim 7.1\,M_{\odot}$.

\citet{lam2022} adopted another approach. In their DW model, in which each data point and its corresponding uncertainty contribute equally to the likelihood, the photometric data dominate. This model leaves large residuals in the astrometric data, especially in the $x$ (east--west) direction.  In their EW model, the photometric and astrometric data sets contribute equally to the likelihood. This model also leaves large residuals, both in astrometric and photometric data. The mass of the lens in the EW model is $2.15^{+0.67}_{-0.54}\,M_{\odot}$, suggesting that the lens may be a neutron star. We judge this model to be unlikely, because it poorly describes both the high-quality photometric OGLE data and the \textit{HST} astrometry. We note that the angle $\varphi$ measured using astrometric data of \citet{lam2022} alone equals to $\varphi=31^{+8}_{-16}$\,deg.

The tension between the direction of the lens-source relative proper motion measured with astrometric and photometric data may have three causes (or some combination of them): (1) systematic errors in the photometric data, (2) systematic errors in the astrometric data, and (3) inadequate model (for example, low-level blending in the \textit{HST} images can alter the inferred amplitude and direction of $\boldsymbol{\theta}_{\rm E} \equiv \thetaE \muhelio/\mu_{\rm rel}$). In this paper, we study the first two possibilities by scrutinizing the available data to search for possible systematic effects. The impact of low-level blending in astrometric microlensing is also discussed.

\begin{figure}
\centering
\includegraphics[width=.5\textwidth]{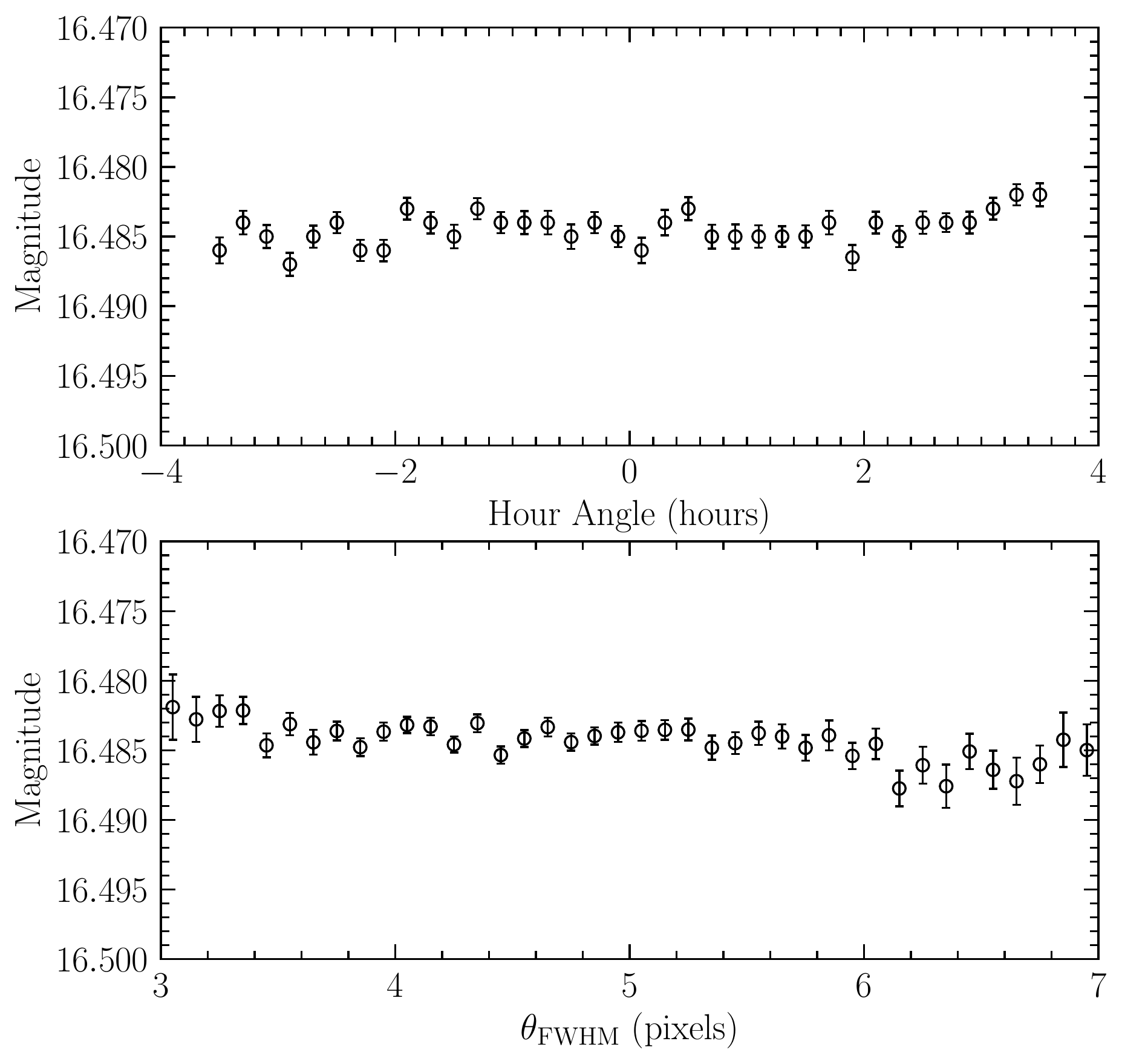}
\caption{Mean OGLE baseline brightness of \OB{} as a function of hour angle (upper panel) and seeing (lower; based on 2013--2016 data, new OGLE reductions). }
\label{fig:ogle_systematics}
\end{figure}

\begin{figure*}
\centering
\includegraphics[width=.48\textwidth]{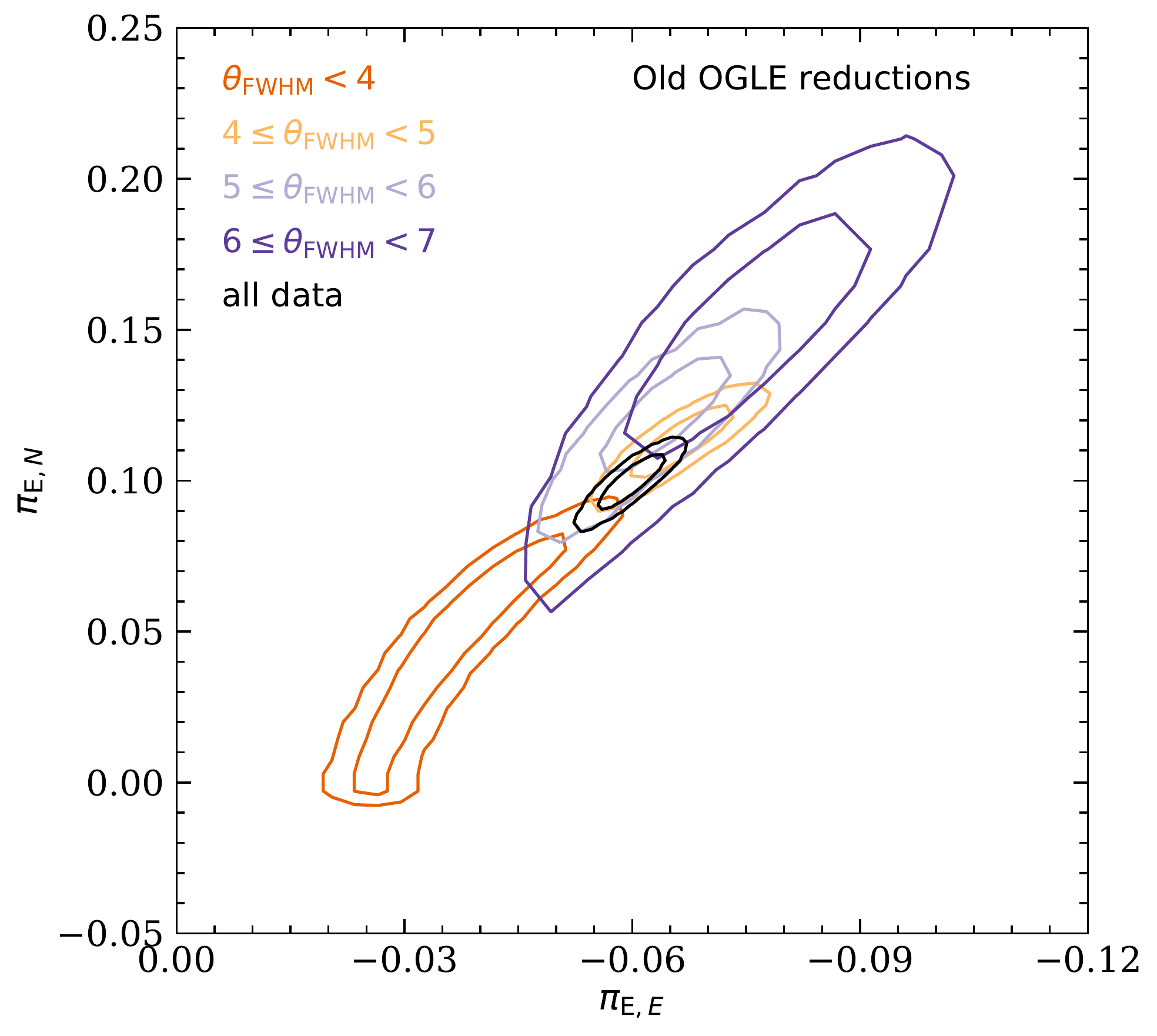}
\includegraphics[width=.48\textwidth]{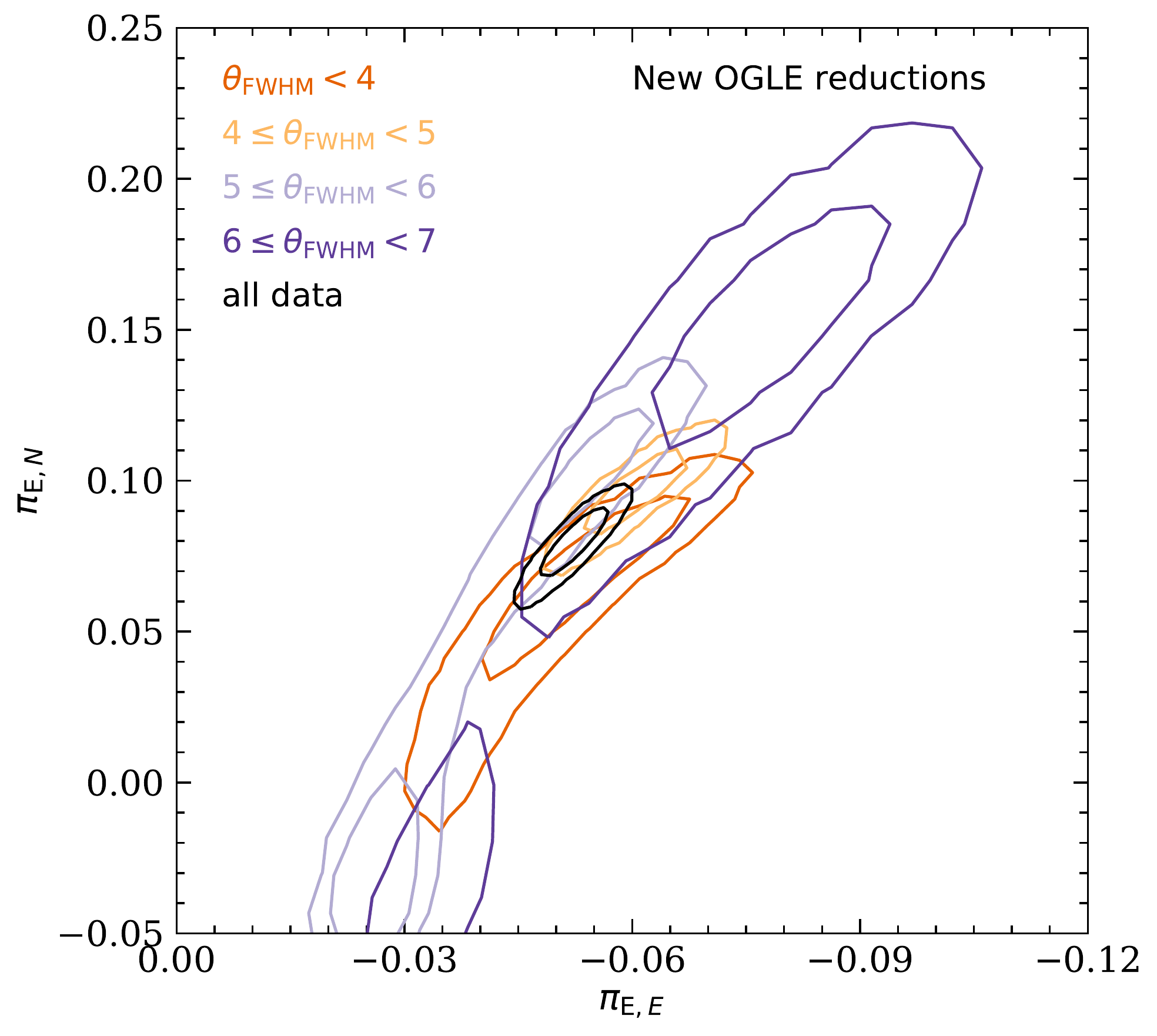}
\caption{Microlensing parallax $\piEN-\piEE$ contours derived using subsets of the OGLE data. Left panel: old reductions. Right panel: new reductions. Both the direction and magnitude of the microlensing parallax vector measured using old reductions when restrict to the excellent seeing data ($\theta_{\rm FWHM} < 4$) are systematically shifted. This behavior is no longer present in the new reductions (right panel). The contours represent 68.3\% and 95.5\% confidence regions.}
\label{fig:ogle_parallax}
\end{figure*}

\section{OGLE-only Model} \label{sec:ogle}

We begin the analysis with the light-curve models based on the OGLE data. Both \citet{sahu2022} and \citet{lam2022} make use of different ground-based photometric data sets. However, to minimize the systematic effects, it is essential to restrict the ground data sets to the most reliable ones (or their subsets). Whenever possible, this means restricting the data set to OGLE, which has shown the greatest stability in many events over more than two decades. Moreover, even with this restriction, it will be necessary to carefully vet for systematic effects that could affect the model. OGLE photometry is calculated using the difference image analysis method \citep{tomaney1996,alard1998} as implemented by \citet{wozniak2000}.

In this paper, we consider two OGLE data sets. One is identical to the one employed by \citet{sahu2022} and \citet{lam2022} and is based on a reference image that is used in real-time reductions at the telescope. That image (like all reference images used in real-time OGLE-IV reductions) is the average of several images. Its components were taken in 2010. The full width at half maximum of the reference frame PSF is $\theta_{\rm FWHM}=3.5$ ($\theta_{\rm FWHM}$ is expressed in pixels, one pixel is $0.26''$). As we show below, the microlensing parallax measured using data collected in exceptionally good seeing conditions ($\theta_{\rm FWHM} \lesssim 4$) is different from that measured using the remaining data. To check for possible systematic effects related to image subtractions (especially when the science image PSF is better than that of the reference frame), we create a new reference image with the best possible $\theta_{\rm FWHM}=2.95$ (based on individual best seeing images collected in 2015, i.e., at the event baseline), recreate the subtracted images using the new reference frame, and recalculate the photometry of the event. The data sets based on different reference images are called \textit{old} and \textit{new} OGLE reductions, respectively.

First, we correct the photometric uncertainties using the method described by \citet{skowron2016}. The uncertainties reported by the OGLE photometric pipeline are known to be underestimated, and \citet{skowron2016} provided a procedure that can be used to correct the uncertainties so that they reflect the actual observed scatter at a given magnitude and for a given detector of the OGLE-IV camera. The corrected uncertainties adequately describe the scatter in the data, as evidenced by the reduced $\chi^2$ statistic being close to unity (Table~\ref{tab:fit}). The median uncertainty is 13 and 11\,mmag in old and new OGLE reductions, respectively.

Next, we search for possible systematic errors in the OGLE data. The source of OGLE-2011-BLG-0462 is heavily blended with a brighter (unresolved in ground-based images) star and a few fainter ones (see Figure~1 of \citealt{lam2022} or Figure~4 of \citealt{sahu2022}). Thus, one might expect some systematic errors. Nevertheless, our investigation of the 10 yr OGLE light curve does not reveal any annual- or air-mass-related trends that could be caused by, e.g., differential refraction (as shown in the upper panel of Figure~\ref{fig:ogle_systematics}). We also examine light curves of neighbor stars of similar brightness. We do not find any evidence for systematic errors with amplitude of 0.01\,mag or higher.

We fit the point-source point-lens model with microlensing parallax to the available data. The magnifications are calculated in the geocentric frame \citep{gould2004} that is better adapted to testing for systematics. The modeling is performed using the Markov Chain Monte Carlo (MCMC) algorithm as coded by \citet{foreman2013}. We assume uniform priors for all parameters of the model.

After initially fitting the data, we examine binned residuals over the whole 10 yr light curve. We find a trend in the data from the first half of 2010. As this was the commissioning time of the OGLE-IV camera, during which many bugs were ironed out of the system, we remove the commissioning data from the first half of 2010 ($\mathrm{HJD}' \equiv \mathrm{HJD}-2450000 < 5376$). We also find a long-duration, low-amplitude bump centered at $\mathrm{HJD}' \sim 8000$ in the old OGLE reductions. This feature is not present in the new reductions. Nevertheless, we remove all points $\mathrm{HJD}'>7700$ because the event is not magnified during these epochs, so they are not germane to the parameters of the event. This enables a strict comparison between the two versions of the data set. Including the later 2017--2019 data in the fits does not significantly change the best-fit parameters of the model.

The results of the fit to the $5376<\mathrm{HJD}'<7700$ data, separately for old and new OGLE reductions, are presented in Table~\ref{tab:fit}. There exists another degenerate solution with negative impact parameter $u_0 \leftrightarrow -u_0$. However, because $u_0 \approx 0$, this solution is nearly identical to the $u_0>0$ model, and, henceforth, we consider only the former (positive $u_0$) solution. In addition to five model parameters ($t_0, \tE, u_0, \piEN, \piEE$), we report a number of derived parameters, including the effective timescale $t_{\rm eff}=\tE |u_0|$, microlensing parallax $\piE$, position angle of the relative lens-source proper-motion vector $\varphi$, blend-to-source flux ratio $(\Fb/\Fs)_{\rm OGLE}$, and source magnitude $I_{S}$. Here $\Fs$ is the flux of the source, and $\Fb$ is the sum of fluxes of blends (which may be from the lens, companions to the lens and source, and/or unrelated neighbor stars).

We then investigate whether seeing variation may induce low-level systematic effects in the OGLE photometry of OGLE-2011-BLG-0462. We find a small, at most 5~mmag trend in the baseline magnitude at very good ($\theta_{\rm FWHM} < 4$) or very poor seeing ($\theta_{\rm FWHM} > 7$) in the old OGLE reductions. These systematics are no longer present in the new reductions, however (lower panel of Figure~\ref{fig:ogle_systematics}).

We also check how seeing affects photometric models of OGLE-2011-BLG-0462. The OGLE light curve is divided into four subsamples based on ranges of seeing ($\theta_{\rm FWHM}<4$, $4 \leq \theta_{\rm FWHM}<5$, $5 \leq \theta_{\rm FWHM}<6$, and $6 \leq \theta_{\rm FWHM}<7$). Figure~\ref{fig:ogle_parallax} shows posterior 68.3\% and 95.5\% confidence regions in the $\piEN-\piEE$ plane derived using subsets of the OGLE data, for old (left panel) and new (right panel) reductions, respectively. The left panel shows that the microlensing parallax measured using excellent seeing data ($\theta_{\rm FWHM}<4$) differs from that measured in other seeing ranges. Both the magnitude and direction of the parallax vector are different. Constraints derived from data from the seeing ranges $4 \leq \theta_{\rm FWHM} < 5$, $5 \leq \theta_{\rm FWHM} < 6$, and $6 \leq \theta_{\rm FWHM} < 7$ overlap each other, while those found using excellent seeing data favor a smaller parallax ($0.060 \pm 0.027$ versus $0.129 \pm 0.008$ for $4 \leq \theta_{\rm FWHM} < 7$ data) and a smaller position angle ($337.0 \pm 20.0$ deg versus $357.8 \pm 2.2$ deg).

On the other hand, the right panel of Figure~\ref{fig:ogle_parallax} shows the constraints on the microlensing parallax measured using subsets of new OGLE reductions. All contours now overlap. The microlensing parallax calculated using all available data equals to $\piE = 0.095 \pm 0.009$ and is $\approx 18\%$ lower than that measured using the old reductions. Moreover, the direction of the lens-source relative proper motion calculated using new reductions ($345.1 \pm 3.7$\,deg) is consistent with that measured with the \textit{HST} astrometry only \citep{sahu2022}, as shown in Figure~\ref{fig:pm1}. All these facts demonstrate that the old reductions are affected by systematic errors resulting from imperfect image subtractions with larger reference image PSF. The most likely reason for this systematics is a bright neighbor star located only 1.5\,pixels from the source. In the rest of the paper, we use only new OGLE reductions in the final models. We use all data collected between $5376<\mathrm{HJD}'<7700$ without any other cuts.

\begin{figure}
\includegraphics[width=.5\textwidth]{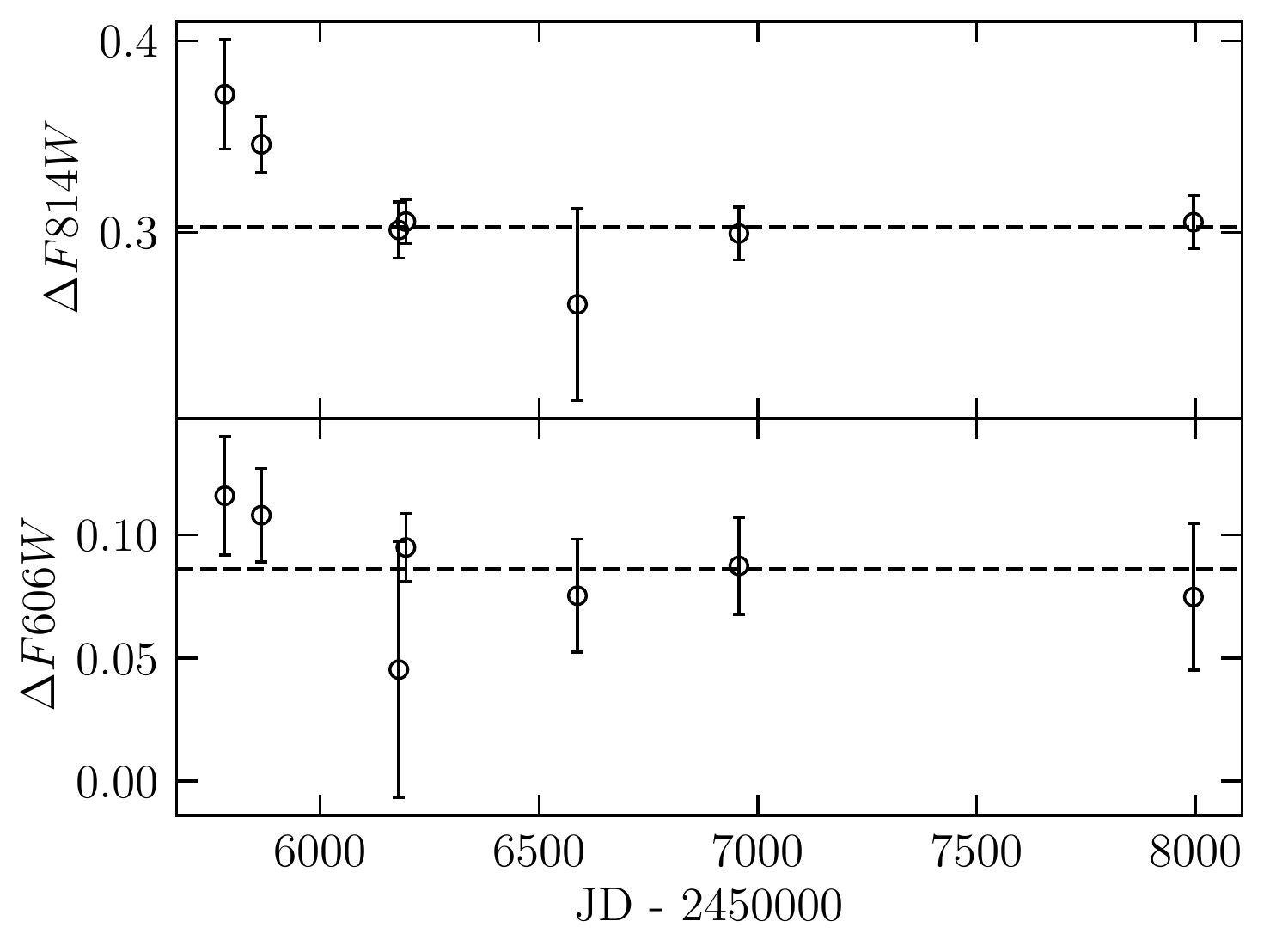}
\caption{Difference between the $F814W$ and $F606W$ photometric measurements of \OB{} taken by \citet{lam2022} and \citet{sahu2022}. The dashed lines indicate a constant shift of $0.303 \pm 0.007$ and $0.086 \pm 0.010$~mag in $F814W$ and $F606W$, respectively.}
\label{fig:hst_mag}
\end{figure}

\begin{figure*}
\includegraphics[width=.5\textwidth]{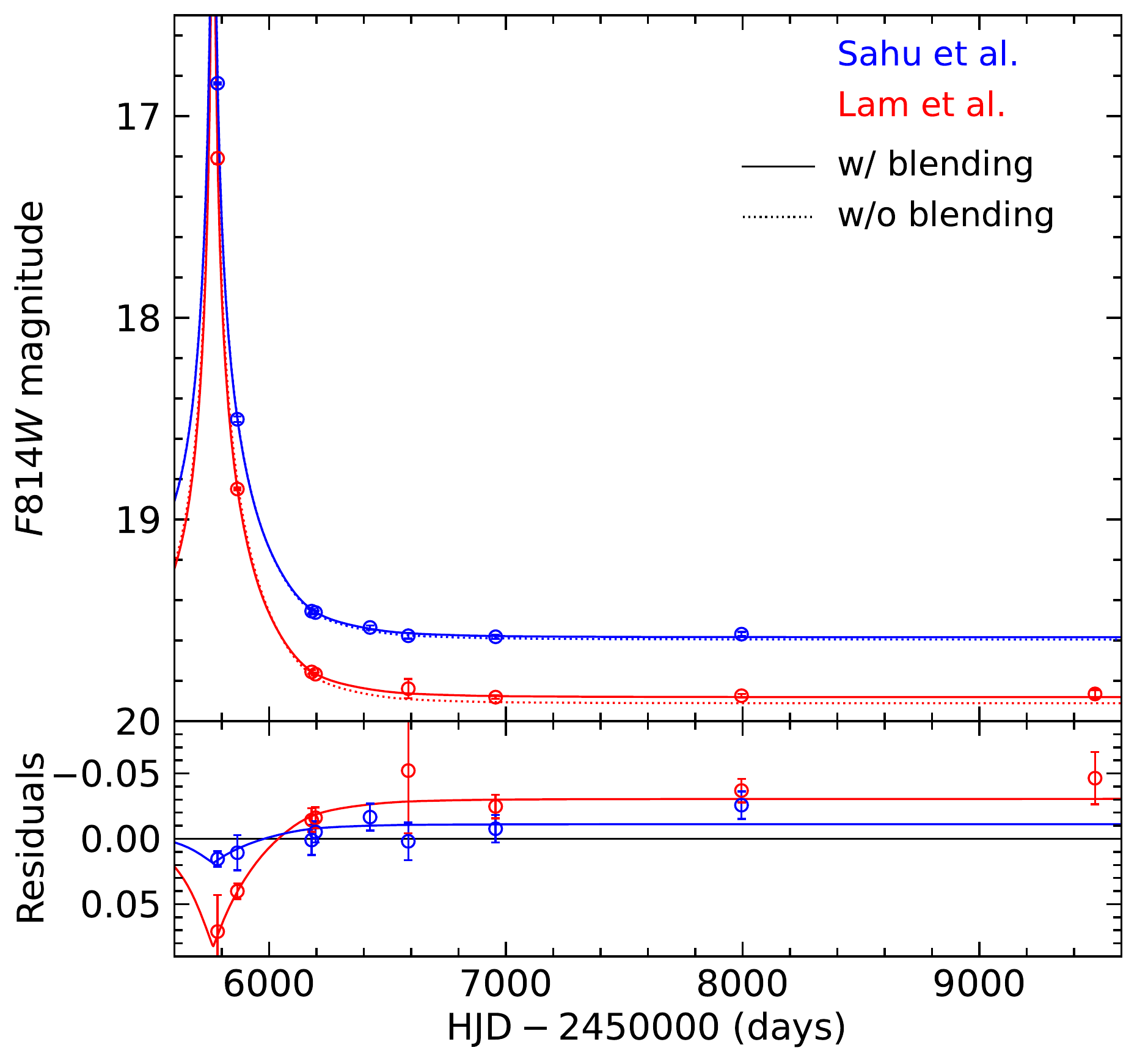}
\includegraphics[width=.5\textwidth]{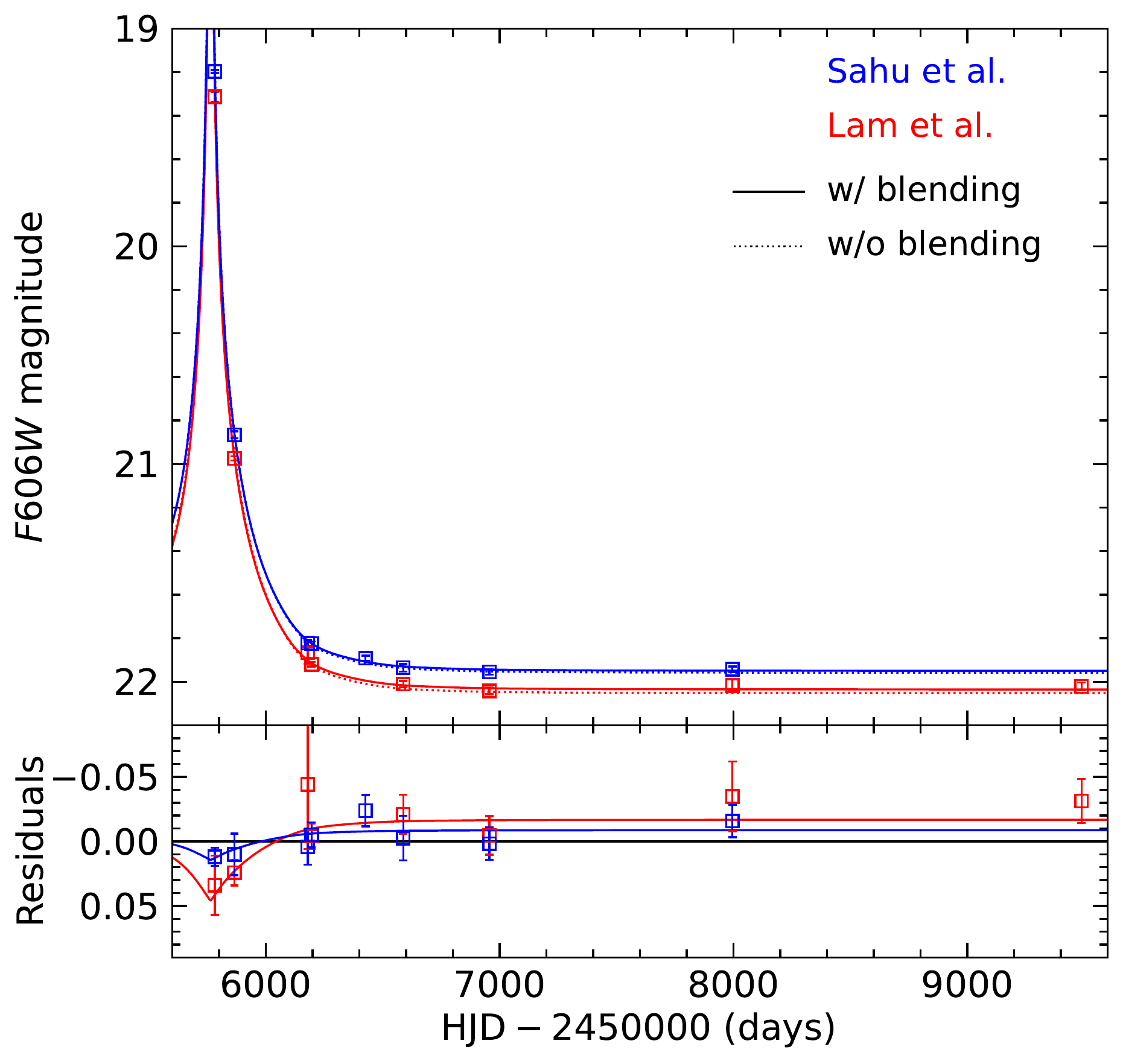}
\caption{Fits to the \textit{HST} photometric data in the $F814W$ (left panel) and $F606W$ (right) filters based on the best-fitting model to the photometric OGLE data only (Section~\ref{sec:ogle}). \citet{sahu2022} and \citet{lam2022} photometric reductions are marked with blue and red colors, respectively. Solid (dotted) lines mark best-fitting models with (without) blending. Lower panels show residuals from the best-fitting models without blending.}
\label{fig:hst_blending}
\end{figure*}

\section{Blending in the \textit{HST} Data} \label{sec:seeing}

Figure~\ref{fig:hst_mag} shows the difference between the $F814W$ and $F606W$ photometric measurements of \OB{} taken by \citet{lam2022} and \citet{sahu2022}. In addition to a constant shift of $0.303 \pm 0.007$ and $0.086 \pm 0.010$~mag in $F814W$ and $F606W$, respectively, the first two epochs are systematically shifted by 0.069 and 0.043~mag ($F814W$) and 0.030 and 0.022~mag ($F606W$), which exceeds their error bars, substantially so in the case of $F814W$. The event was highly magnified during these two epochs (magnifications of $12.87 \pm 0.13$ and $2.76 \pm 0.03$, respectively), indicating that low-level systematics ($2\%-7\%$ of flux) exist in at least one of the \textit{HST} data reductions.

These systematics are reflected in the measurements of the blend flux. Including the \textit{HST} photometric data in the fits does not influence the best-fit parameters due to the small number of \textit{HST} epochs. Data from these two \textit{HST} epochs, when the event was magnified, mainly constrain the source and blend fluxes with virtually no impact on the inferred parameters of the microlensing model. Thus, we use the posterior distributions derived using only OGLE data (Section~\ref{sec:ogle}) to measure the \textit{HST} source and blend fluxes. Because the observed flux depends linearly on $\Fs$ and $\Fb$, the flux parameters are calculated analytically using the least-squared method for each set of $(t_0, \tE, u_0, \piEN, \piEE)$ from the MCMC chain.

The results are summarized in Table~\ref{tab:fit}, separately for the \citet{sahu2022} and \citet{lam2022} reductions. Blending in \citet{sahu2022} reductions is small in an absolute sense, but formally inconsistent with zero, i.e., $\Delta\chi^2=18.6$ for 2~degrees of freedom (dof), i.e., $(\Fb/\Fs)_{\mathrm{HST},F814W}=0.028 \pm 0.011$ and $(\Fb/\Fs)_{\mathrm{HST},F606W}=0.022 \pm 0.011$. Moreover, \citet{lam2022} reductions indicate much stronger blending, with $\Delta\chi^2 = 101.0$, $(\Fb/\Fs)_{\mathrm{HST},F814W}=0.110 \pm 0.014$ and $(\Fb/\Fs)_{\mathrm{HST},F606W}=0.059 \pm 0.013$. Figure~\ref{fig:hst_blending} shows the best-fitting models to the \textit{HST} photometric data.

The chance of significant blending from random interlopers is small ($<1\%$), so that we do not expect detectable blending except from a possible companion to the lens or source.
After cleaning the OGLE data, we find that the blending is 3.9 times closer to zero for Sahu et al.~reductions than Lam et al.~reductions. The remaining difference could be explained either by small remaining systematic errors in either OGLE or \textit{HST} photometry (and zero blending because no luminous star) or by a faint blend, i.e., 3.9~mag below the source, which could easily be an early M dwarf companion to the source, which are very common. High-resolution spectroscopic observations of the event by \citet{bensby2013} indicate that the source star is a late G-type subgiant with a mass of about $1.18\,M_{\odot}$. The multiplicity of solar-type stars (with spectral types $\sim$F6--K3) was studied by \citet{raghavan2010} who found that the distribution of binary mass ratios is roughly flat in the range 0.2--0.95. The orbital period distribution follows a roughly log-normal Gaussian distribution with a maximum at $\sim 300$~yr, which corresponds to the companion separation of $\sim 10$\,mas. Moreover, the color of the blend is $F606W-F814W=2.63^{+0.17}_{-0.08}$\,mag for \citet{sahu2022} reductions and $F606W-F814W=2.80^{+0.11}_{-0.08}$\,mag for \citet{lam2022} reductions. The blend color is consistent with colors of main-sequence stars below the turn-off in the field of the event (see Figure~7 of \citet{sahu2022}).

The blend is unlikely to be a companion to the lens. \citet{sahu2022} did not find any luminous object at the position of the lens at late \textit{HST} epochs, placing an upper limit on its flux of $1\%$ of the source flux. Moreover, a companion to the lens at a projected separation of $0.4 \lesssim a_{\perp} \lesssim 180\,$au would have been easily detectable in the light curve of the event \citep{sahu2022}.

If there is weak blending (we assume here $r=\Fb/\Fs=0.028$) from a source companion that is within the \textit{HST} PSF (at $\Delta\theta < 30$\,mas), then it will displace all astrometric measurements by $\delta\theta(t) \approx r\Delta\theta/(A(t)+r)\approx \epsilon/A(t)$, where $A(t)$ is the magnification and $\epsilon= r\Delta\theta < 0.9\,\mathrm{mas}$. In Appendix~\ref{sec:blending}, we demonstrate that blending can easily account for the remaining offset between the direction of $\muhelio$ measured using photometric and astrometric data, although it cannot account for the whole astrometric deflection that is detected in the \textit{HST} data.

In general, our findings indicate that even a small amount of blending can significantly alter the direction of $\muhelio$ and the angular Einstein radius $\thetaE$ measured using astrometric data. For this case, we have very good OGLE light curve with very well measured parallax, which allows determination of blending by the method of this paper. However, in general, it will be much more difficult to accurately measure blending.

\section{\textit{HST} Astrometric Data} \label{sec:hst}

In the following models to the photometric and astrometric data, we assume that the \textit{HST} source blending can be neglected. This amounts to assuming that the small, but formally significant, blending found above is due to additional low-level systematics in the OGLE and/or \textit{HST} photometric reductions. Note that neither \citet{sahu2022} nor \citet{lam2022} took blending into consideration in their astrometric models.

\begin{figure}
\includegraphics[width=.5\textwidth]{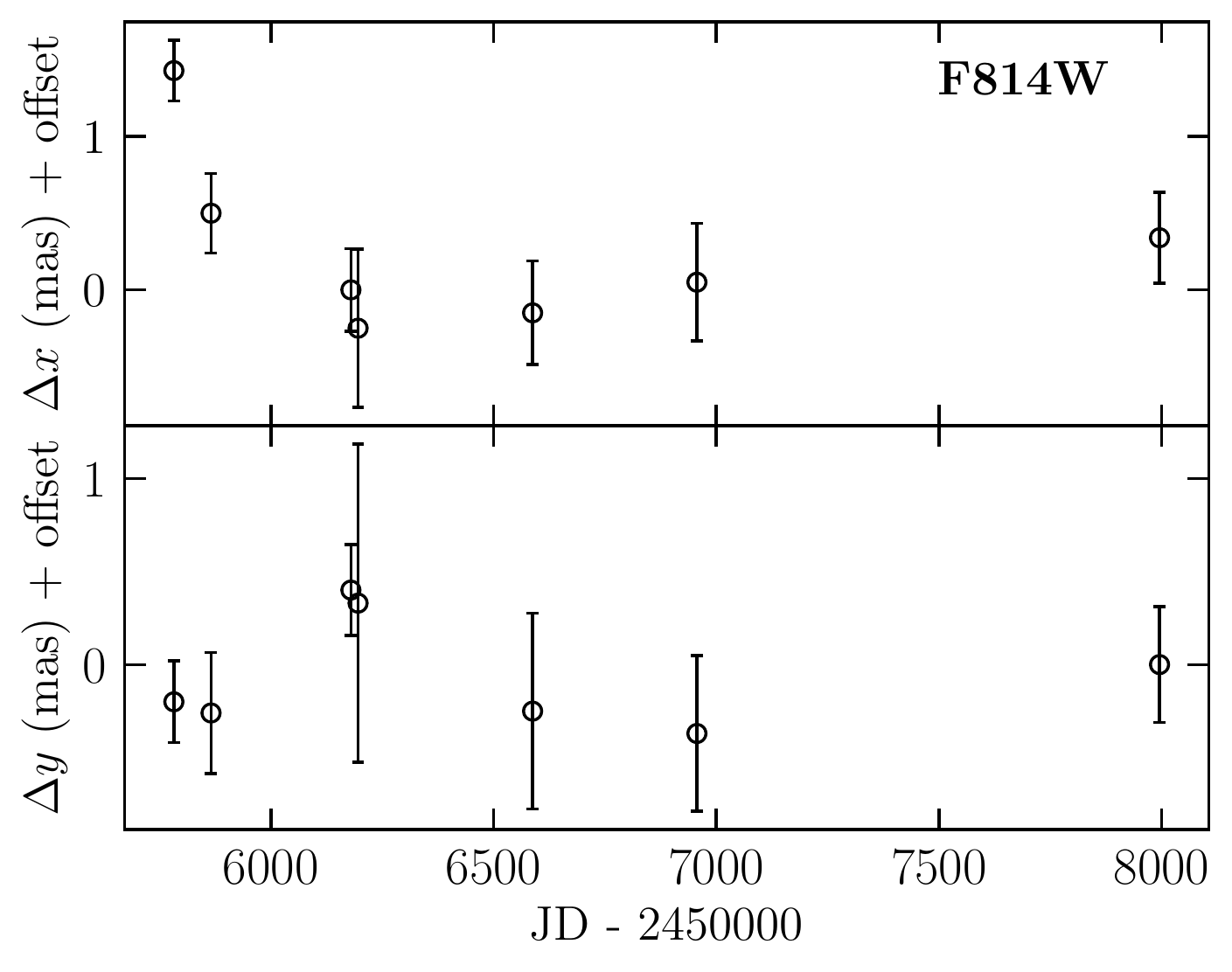}\\
\includegraphics[width=.5\textwidth]{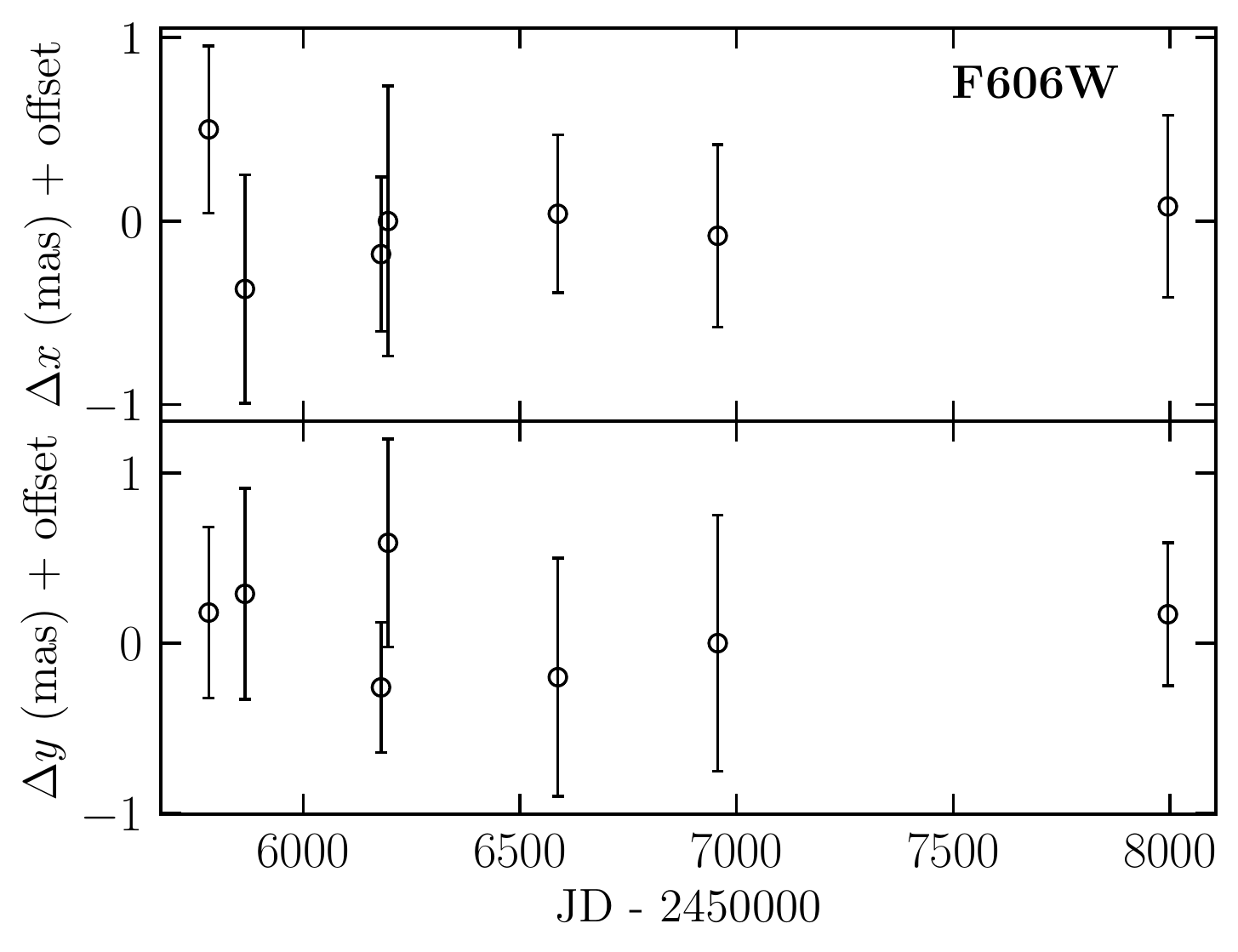}
\caption{Difference between astrometric measurements of \OB{} taken by \citet{lam2022} and \citet{sahu2022} in $F814W$ (upper panel) and $F606W$ filter (lower panel).}
\label{fig:hst_astro}
\end{figure}

The differences between astrometric measurements of \OB{} taken by \citet{lam2022} and those by \citet{sahu2022} are presented in Figure~\ref{fig:hst_astro}, separately for $F814W$ and $F606W$ filters. The $F606W$ data seem to agree well. However, some $F814W$ measurements differ by more than their error bars would indicate. The first two measurements in the $x$ (east--west) direction, taken when the event was magnified, are especially concerning because they differ by $1.43\pm0.20$ and $0.50\pm0.26$\,mas (note that the bright neighbor star is located in the same direction, almost exactly to the west of the source). 

As discussed in Section~\ref{sec:intro}, we are unable to decide which astrometric reductions are better. To quantify the additional uncertainty, we rescale the astrometric uncertainties by adding a constant term in quadrature: $\sigma_{x,i} \rightarrow \sqrt{\sigma_{x,i}^2+\sigma_x^2}$ and $\sigma_{y,i} \rightarrow \sqrt{\sigma_{y,i}^2+\sigma_y^2}$. We treat $\sigma_x$ and $\sigma_y$ as additional parameters of the model, and we find the best-fit parameters by maximizing the following likelihood function \citep{foreman2013}:
\begin{equation}
\ln\mathcal{L} = \ln\mathcal{L}_{\rm phot} + \ln\mathcal{L}_{\mathrm{astr}, x}+ \ln\mathcal{L}_{\mathrm{astr}, y}, 
\end{equation}
where
\begin{align}
\begin{split}
\ln\mathcal{L}_{\rm phot} &= -\frac{1}{2}\sum_i\frac{(F_i-F(t_i))^2}{\sigma^2_{F,i}},\\
\ln\mathcal{L}_{\mathrm{astr}, x} &= -\frac{1}{2}\sum_i\left(\frac{(x_i-x(t_i))^2}{\sigma^2_{x,i}+\sigma^2_x} + \ln (\sigma^2_{x,i}+\sigma^2_x)\right),\\
\ln\mathcal{L}_{\mathrm{astr}, y} &= -\frac{1}{2}\sum_i\left(\frac{(y_i-y(t_i))^2}{\sigma^2_{y,i}+\sigma^2_y} + \ln (\sigma^2_{y,i}+\sigma^2_y)\right).\\
\end{split}
\end{align}

\begin{figure*}
\centering
\includegraphics[width=.42\textwidth]{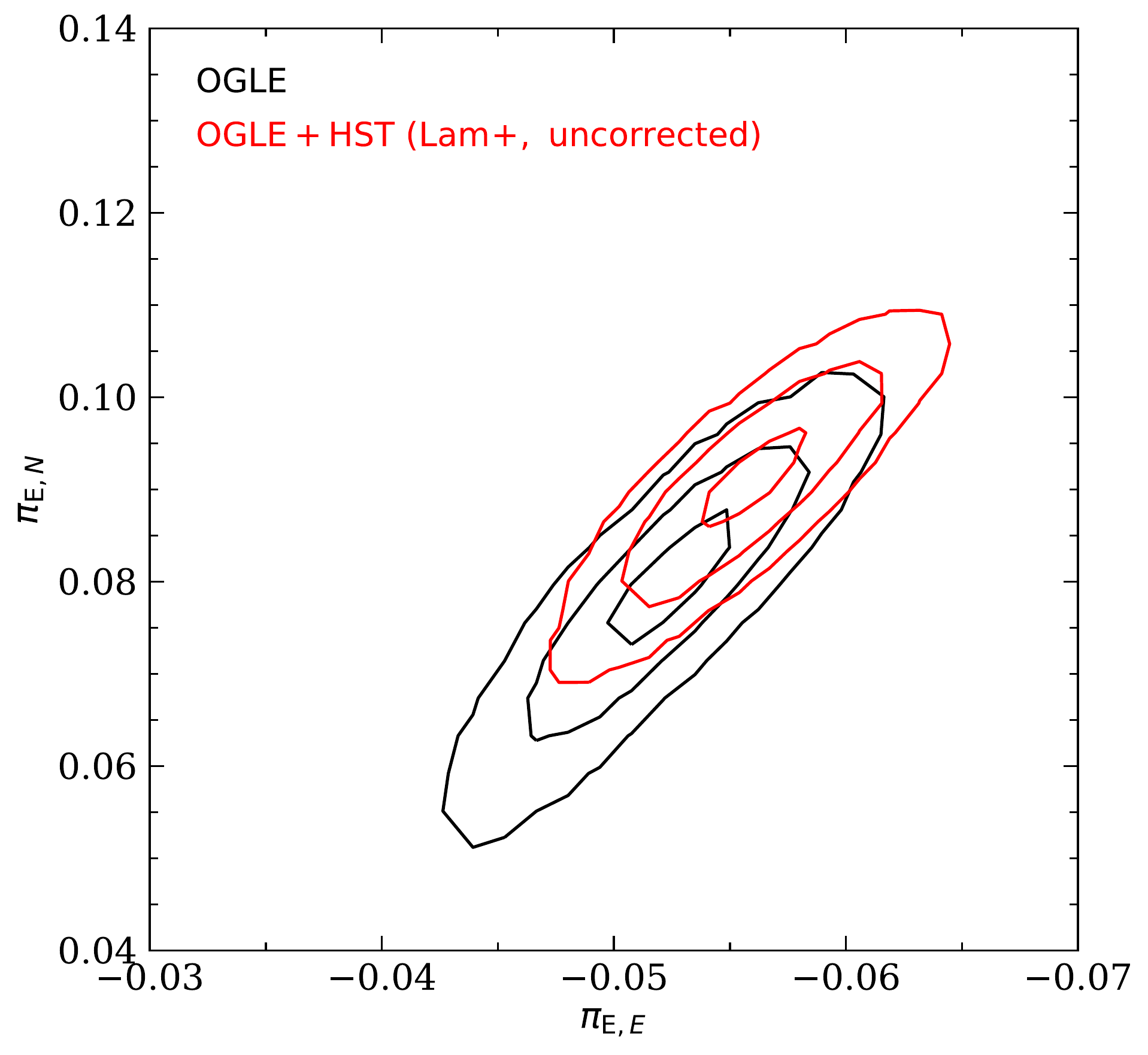}
\includegraphics[width=.42\textwidth]{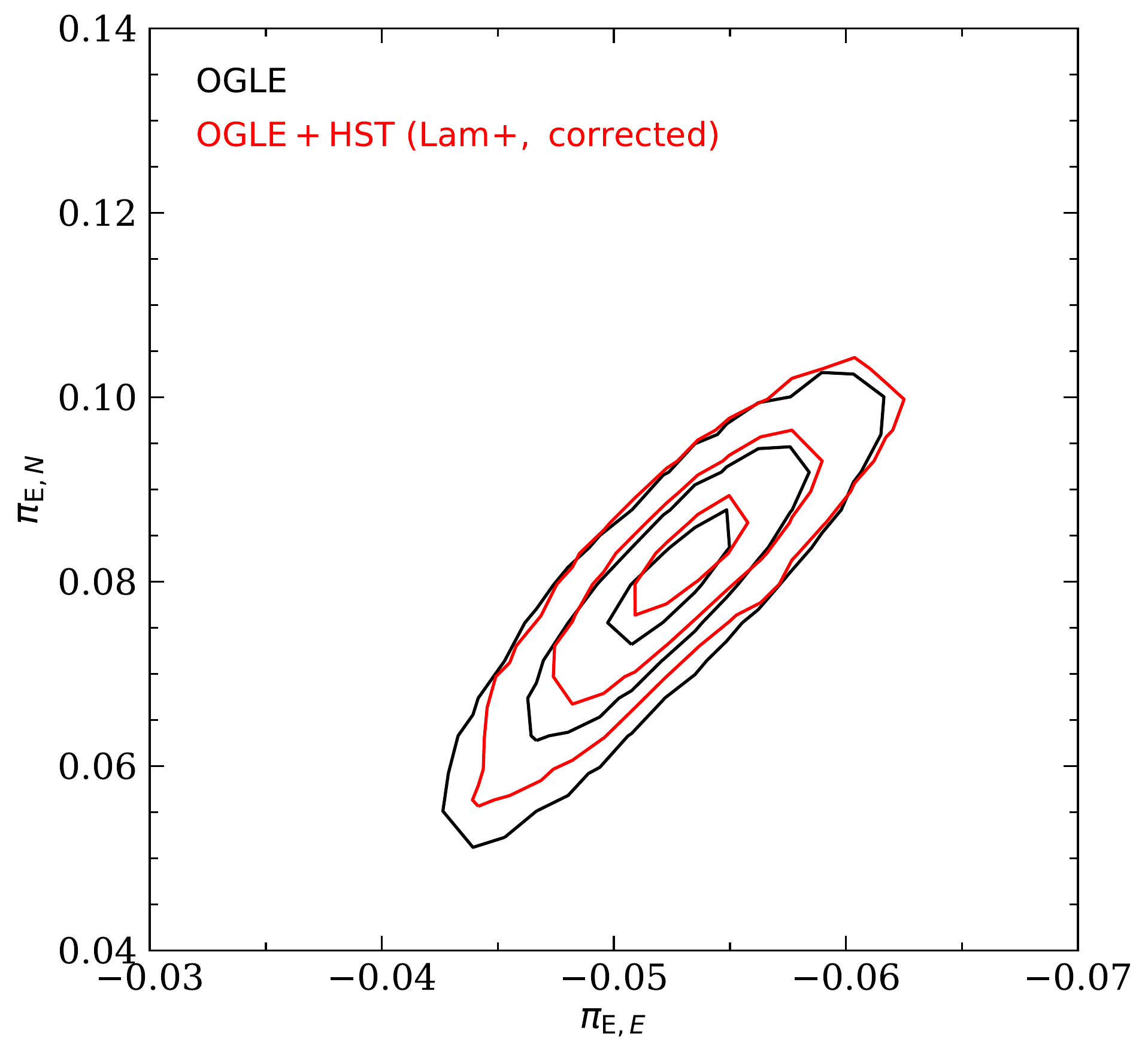}\\
\includegraphics[width=.42\textwidth]{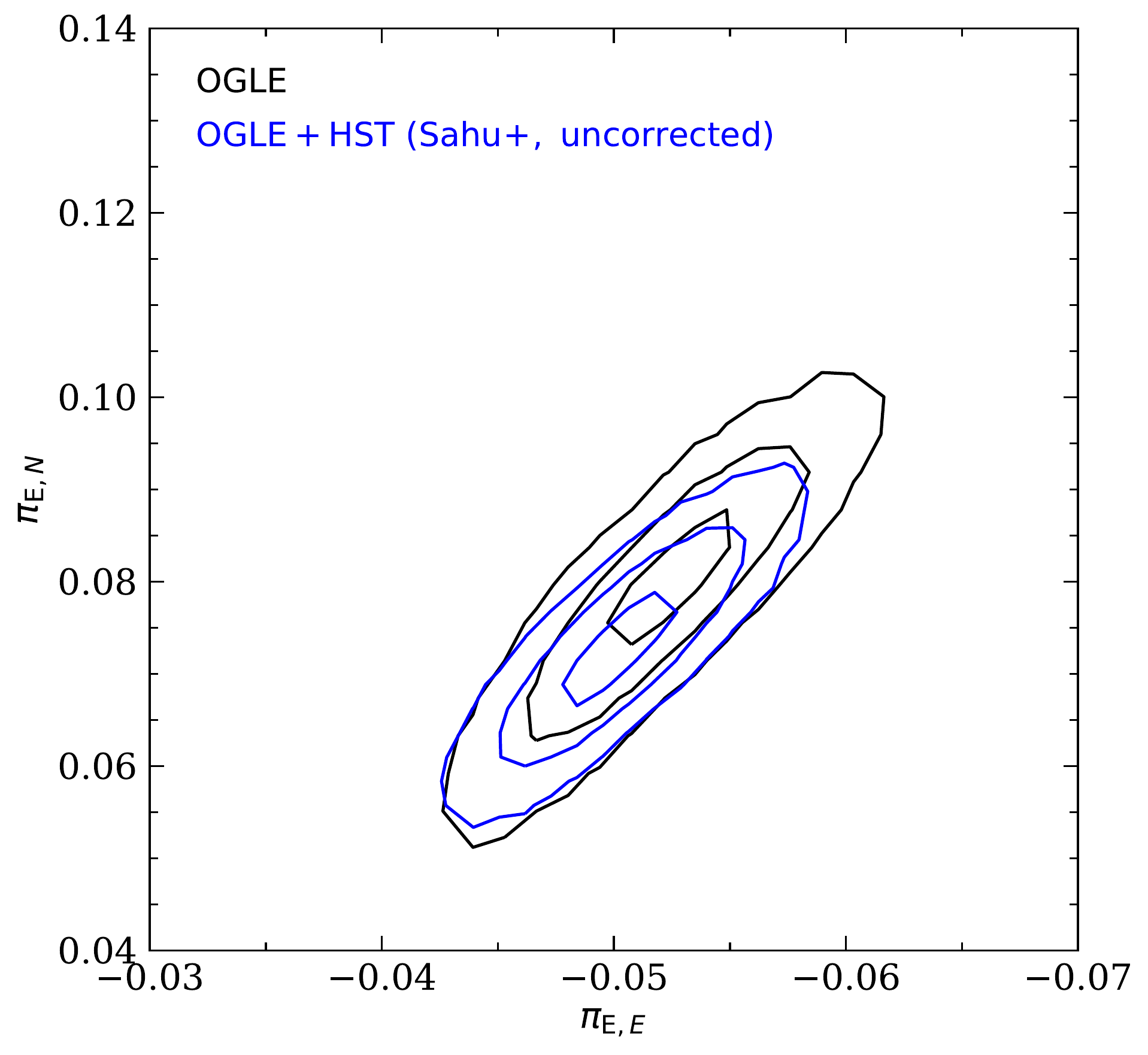}
\includegraphics[width=.42\textwidth]{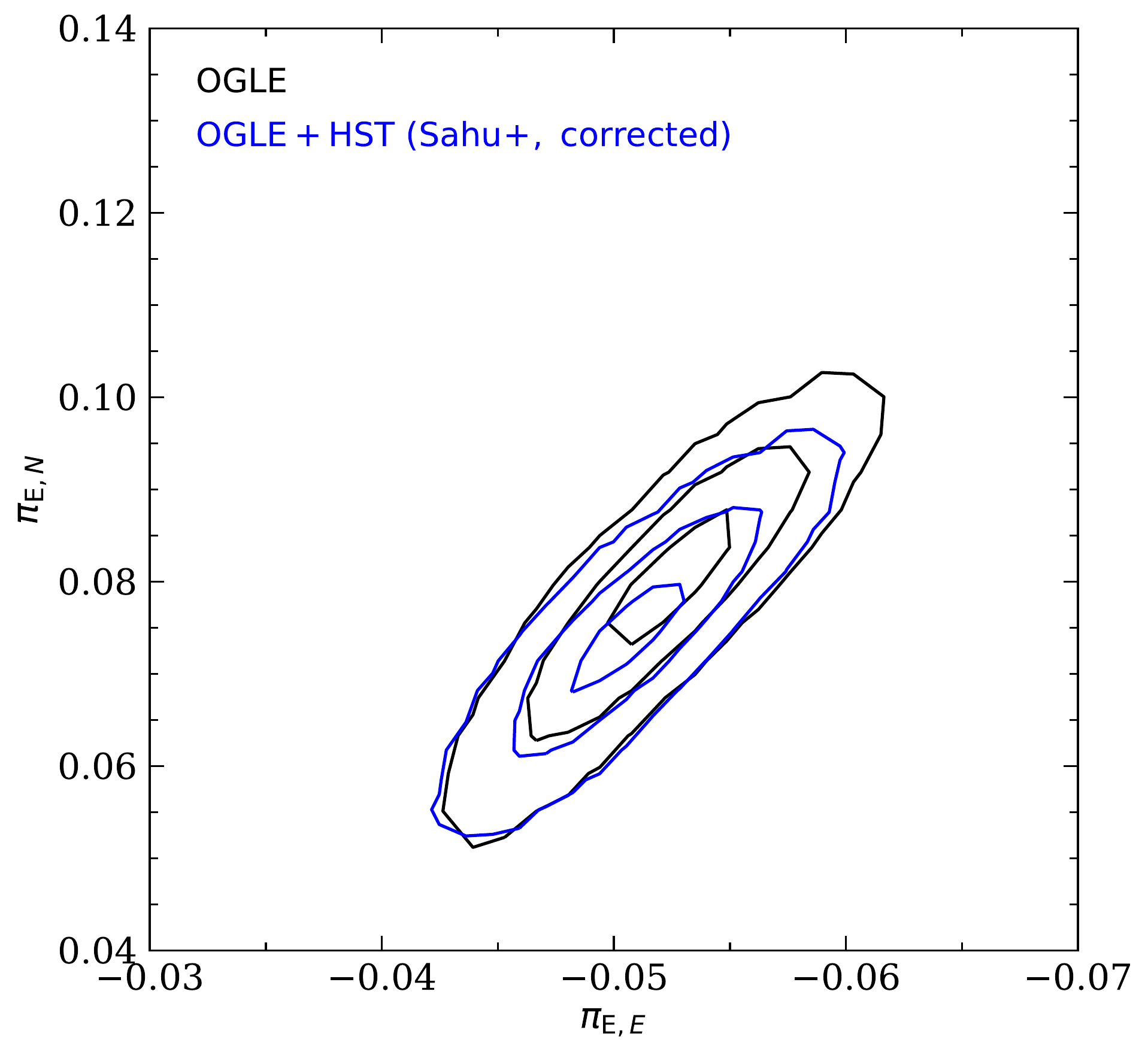}
\caption{Microlensing parallax $\piEN-\piEE$ contours. Black contours are derived using only photometric OGLE data. Red and blue contours are based on a joint OGLE and \textit{HST} fit using \citet{lam2022} and \citet{sahu2022} astrometric reductions, respectively. Contours presented in the left panels are not corrected for possible systematic errors in the astrometric data. The contours represent 39.3\%, 86.5\%, and 98.9\% confidence regions.}
\label{fig:parallax}
\end{figure*}

We assume uniform priors on all parameters of the models. Each photometric and astrometric data point has the same weight (which is the equivalent of the DW model of \citet{lam2022}).
The results of the fits are presented in Table~\ref{tab:fit3}. First, in the second and third columns, we present the best-fit solutions assuming $\sigma_x=0$ and $\sigma_y=0$, that is, no systematic errors in the astrometric data, separately for \citet{sahu2022} and \citet{lam2022} reductions. We are able to reproduce the findings of the original papers, that is, the preferred mass of the lens inferred using \citet{sahu2022} reductions ($7.88 \pm 0.82\,M_{\odot}$) is higher than that found using \citet{lam2022} data ($4.19 \pm 0.77\,M_{\odot}$). These mass measurements differ by nearly $3.3\sigma$. The left panels of Figure~\ref{fig:parallax} present the contours of the microlensing parallax obtained. The black contours mark models based only on the OGLE data, while the red and blue contours are based on joint fits to OGLE photometry and \textit{HST} astrometry. Because of tensions between photometric and astrometric data, especially in the case of \citet{lam2022} astrometric reductions, these contours overlap only partially.

\begin{figure*}
\includegraphics[width=\textwidth]{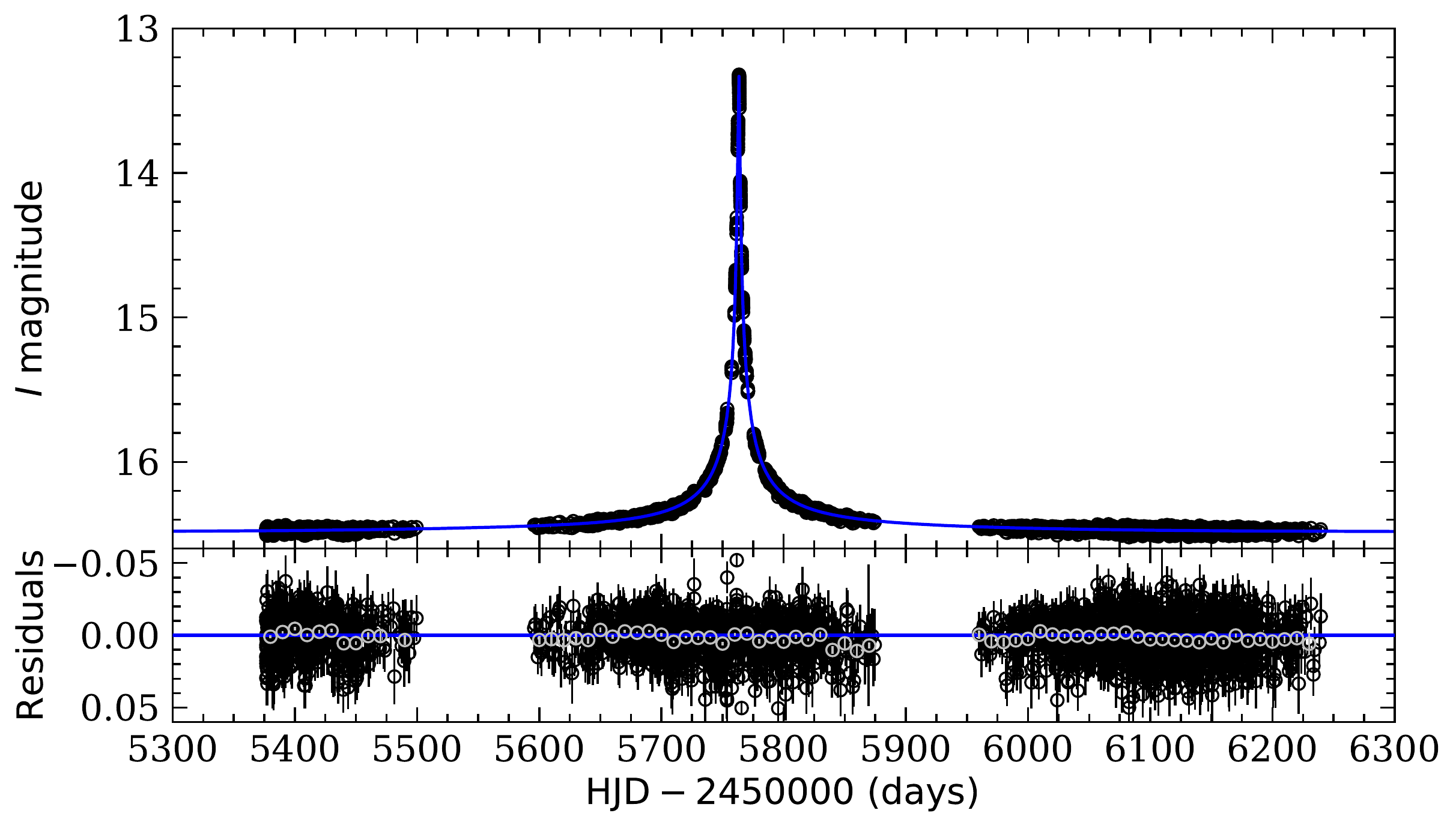}\\
\includegraphics[width=\textwidth]{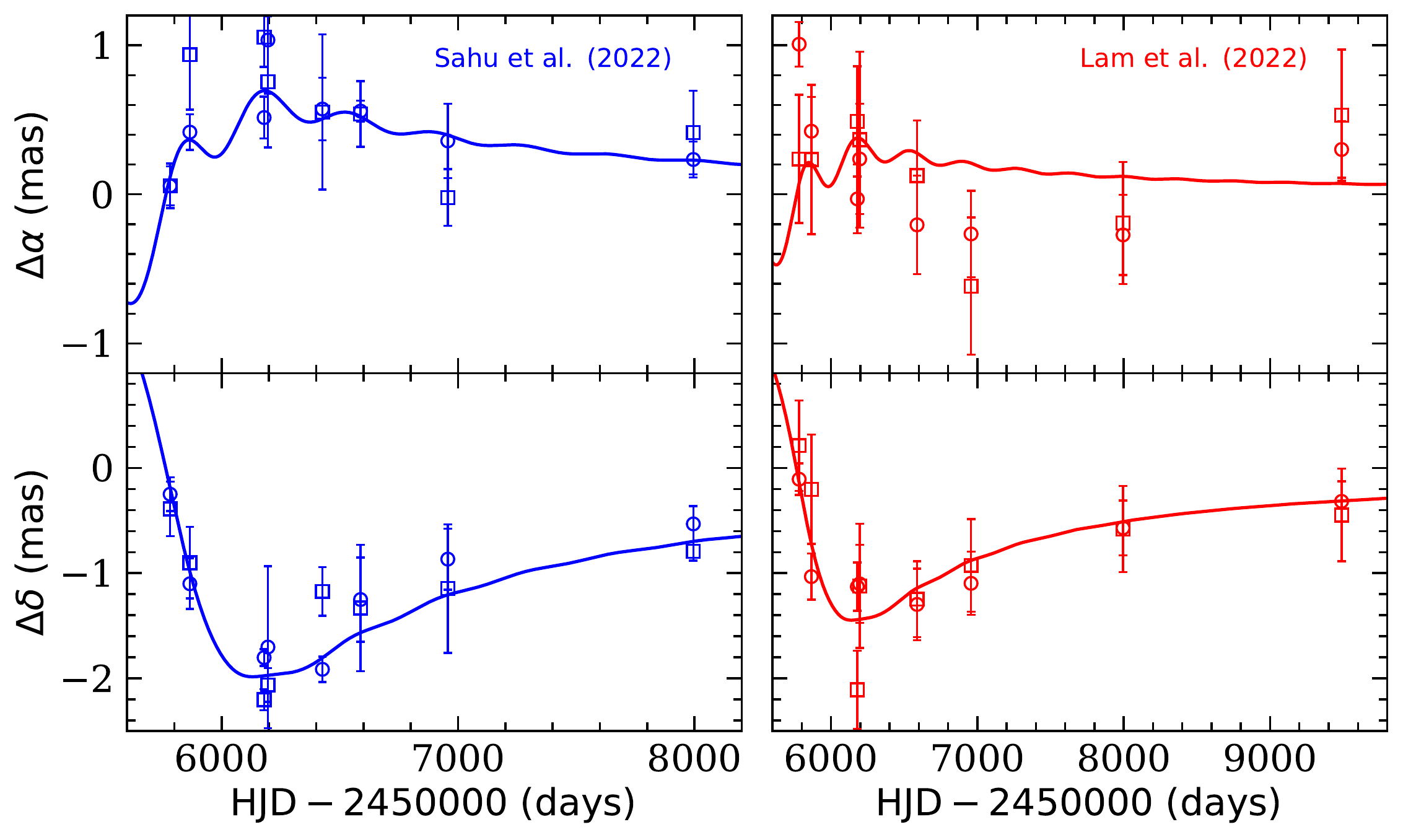}
\caption{Upper panel: photometric OGLE light curve of \OB{} together with a best-fit model to OGLE and \textit{HST} data from \citet{sahu2022}. The best-fit model based on \citet{lam2022} reductions is nearly identical (the difference between these two models is smaller than 1 mmag). Gray data points mark median residuals in bins of 10~days. Lower panel: best-fit model to astrometric data (after subtracting proper motions) reduced by \citet{sahu2022} (blue, left panels) and \citet{lam2022} (red, right panels). $F814W$ data are marked with circles, $F606W$ data wtih squares.}
\label{fig:fits}
\end{figure*}

These tensions are eased when we allow $\sigma_x$ and $\sigma_y$ to vary. The lens mass calculated using \citet{sahu2022} reductions is  slightly lower ($7.80 \pm 1.09\,M_{\odot}$) whereas that calculated using \citet{lam2022} reductions is higher ($5.03 \pm 1.10\,M_{\odot}$). The masses are still discrepant, but the difference is at a $1.8\sigma$ level. Similarly, the constraints on the microlensing parallax (presented in the right panels of Figure~\ref{fig:parallax}) agree well with those calculated using only OGLE data. Figure~\ref{fig:fits} shows residuals from the best-fit models.

The inferred $\sigma_x$ and $\sigma_y$ quantify possible systematic errors in the astrometric data. For \citet{lam2022} reductions, we obtain $\sigma_x=0.433 ^{+0.138}_{-0.104}$\,mas, and $\sigma_y=0.124 ^{+0.130}_{-0.088}$\,mas. Indeed, Figure~\ref{fig:fits} and figure~2 of \citet{lam2022} show that the astrometric residuals in the $x$ direction are much larger than those in the $y$ direction. The systematics errors in the $x$ and $y$ direction are of the order of $\sim0.01$ and $\sim 0.003$ pixel size of the \textit{HST} WFC3 camera, respectively. On the other hand, for \citet{sahu2022} reductions, the first parameter is smaller ($\sigma_x=0.088 ^{+0.088}_{-0.061}$\,mas) whereas the second ($\sigma_y=0.189 ^{+0.092}_{-0.072}$\,mas) is similar to that found for \citet{lam2022} reductions.

We also fit a model with an alternative weighting scheme, which is equivalent to the EW model of \citet{lam2022}. In that model, the direction of the microlensing parallax vector is constrained mostly by the astrometric data ($\varphi = 338.3 \pm 0.2$\,deg for Sahu reductions and $\varphi=27.3 \pm 0.4$\,deg for Lam reductions). The direction of the best-fitting parallax vector found for Lam astrometric reductions is inconsistent with that found from the photometry, which is reflected by large residuals from the light-curve model with an amplitude of $\sim 0.02$\,mag (similar to those presented in figure~8 of \citealt{lam2022_supl}), while we have shown that systematics in the OGLE data are at least an order of magnitude smaller than that (Figure~\ref{fig:ogle_systematics}).

\section{Discussion and Summary}

As shown in the previous sections, using new photometric OGLE reductions reduces the tension between astrometric and photometric measurements. However, different \textit{HST} data reductions yield different lens masses. Figure~\ref{fig:hst_astro} demonstrates that at least one of the astrometric data sets suffers from unknown systematic errors. Including the systematic error term in the astrometric model improves the fit by $\Delta\chi^2=3$ for \citet{sahu2022} reductions, which is a marginal improvement. (Note that, by adding two free parameters to the model, we expect some $\Delta\chi^2=2$ improvement; see \citealt{gould2003}.) However, in models based on \citet{lam2022} reductions, adding the systematic astrometric error terms improves the fit by $\Delta\chi^2=20$, which is statistically significant. This indicates there is no strong evidence for systematic errors in the \textit{HST} data reductions by \citet{sahu2022}.

Moreover, in the model based on astrometric data of \citet{lam2022} only, the direction of the lens-source relative proper motion equals to $\varphi=31^{+8}_{-16}$\,deg and is different from that measured using OGLE data only ($\varphi = 345.1 \pm 3.7$\,deg) and astrometric data of \citet{sahu2022} only ($\varphi=337.9 \pm 5.0$\,deg). This indicates that \citet{lam2022} reductions are affected by systematic errors. This conclusion holds true even if we repeat the model fitting without an extra \textit{HST} epoch from 2021 that appears to change the direction of the baseline source proper motion. We hope that both teams analyzing the \textit{HST} data work together to track down the origin of systematic errors. 

In our preferred model, based on \citet{sahu2022} astrometric data and without the systematic astrometric error term, the lens mass and distance are as follows:
\begin{align}
\begin{split}
M &= 7.88 \pm 0.82\,M_{\odot},\\
D_l &= 1.49 \pm 0.12\,\mathrm{kpc},
\end{split}
\end{align}
assuming the distance to the source of $D_s = 5.9 \pm 1.3$\,kpc \citep{sahu2022}.
A main-sequence star of that mass located at that distance would have been easily detectable in the \textit{HST} images \citep{sahu2022}, so the lens must be an isolated black hole.

Table~\ref{tab:fit3} also presents the inferred proper motion of the lens and its space velocity calculated by assuming the velocity of the Sun with respect to the local standard of rest from \citet{schonrich2010}. The transverse velocity of $40.1 \pm 3.2\,\mathrm{km\,s}^{-1}$ could be due to a natal kick that was received by the black hole at birth \citep{andrews2022}.

\citet{sahu2022} measured the distance to the source by fitting theoretical isochrones to the location of the source on the $\log g$ vs.~effective temperature diagram using stellar parameters from \citet{bensby2013}. \citet{bensby2013} also reported the absolute magnitude of the source of $M_I = 2.83 \pm 0.04$\,mag, which favors a larger distance to the source. By comparing the position of the source relative to the red clump centroid in the color--magnitude diagram, we find the source distance of $D_s = 8.8 \pm 1.4$\,kpc. The corresponding distance and transverse velocity of the lens are $1.62 \pm 0.15$\,kpc and $43.4 \pm 3.8\,\mathrm{km\,s}^{-1}$, respectively.

The case of \OB{} is another proof that nature guards its secrets zealously. Blending from a close bright neighbor is the main culprit: the bright neighbor has induced systematic errors both in the astrometric and photometric measurements. This is an extreme case of blending. In most cases, detections of dark stellar remnants with gravitational microlensing should be more accurate.

\section*{Acknowledgements}

We thank Casey Lam, Jessica Lu, and Kailash Sahu for fruitful discussions about \OB{} and Radek Poleski for reading the manuscript.
This research was funded in part by National Science Centre, Poland, grant OPUS 2021/41/B/ST9/00252 awarded to P.M. 

\bibliographystyle{aasjournal}
\bibliography{pap}

\begin{thebibliography}{}
\expandafter\ifx\csname natexlab\endcsname\relax\def\natexlab#1{#1}\fi

\bibitem[{{Alard} \& {Lupton}(1998)}]{alard1998}
{Alard}, C., \& {Lupton}, R.~H. 1998, \apj, 503, 325

\bibitem[{{Andrews} \& {Kalogera}(2022)}]{andrews2022}
{Andrews}, J.~J., \& {Kalogera}, V. 2022, \apj, 930, 159

\bibitem[{{Bensby} {et~al.}(2013){Bensby}, {Yee}, {Feltzing}, {Johnson},
  {Gould}, {Cohen}, {Asplund}, {Mel{\'e}ndez}, {Lucatello}, {Han}, {Thompson},
  {Gal-Yam}, {Udalski}, {Bennett}, {Bond}, {Kohei}, {Sumi}, {Suzuki}, {Suzuki},
  {Takino}, {Tristram}, {Yamai}, \& {Yonehara}}]{bensby2013}
{Bensby}, T., {Yee}, J.~C., {Feltzing}, S., {et~al.} 2013, \aap, 549, A147

\bibitem[{{Delplancke} {et~al.}(2001){Delplancke}, {G{\'o}rski}, \&
  {Richichi}}]{delplancke2001}
{Delplancke}, F., {G{\'o}rski}, K.~M., \& {Richichi}, A. 2001, \aap, 375, 701

\bibitem[{{Foreman-Mackey} {et~al.}(2013){Foreman-Mackey}, {Hogg}, {Lang}, \&
  {Goodman}}]{foreman2013}
{Foreman-Mackey}, D., {Hogg}, D.~W., {Lang}, D., \& {Goodman}, J. 2013, \pasp,
  125, 306

\bibitem[{{Gould}(1994)}]{gould1994}
{Gould}, A. 1994, \apjl, 421, L71

\bibitem[{{Gould}(2003)}]{gould2003}
{Gould}, A. 2003, arXiv e-prints, arXiv:0310577

\bibitem[{{Gould}(2004)}]{gould2004}
{Gould}, A. 2004, \apj, 606, 319

\bibitem[{{Hog} {et~al.}(1995){Hog}, {Novikov}, \& {Polnarev}}]{hog1995}
{Hog}, E., {Novikov}, I.~D., \& {Polnarev}, A.~G. 1995, \aap, 294, 287

\bibitem[{{Lam} {et~al.}(2022{\natexlab{a}}){Lam}, {Lu}, {Udalski}, {Bond},
  {Bennett}, {Skowron}, {Mroz}, {Poleski}, {Sumi}, {Szymanski}, {Kozlowski},
  {Pietrukowicz}, {Soszynski}, {Ulaczyk}, {Wyrzykowski}, {Miyazaki}, {Suzuki},
  {Koshimoto}, {Rattenbury}, {Hosek}, {Abe}, {Barry}, {Bhattacharya}, {Fukui},
  {Fujii}, {Hirao}, {Itow}, {Kirikawa}, {Kondo}, {Matsubara}, {Matsumoto},
  {Muraki}, {Olmschenk}, {Ranc}, {Okamura}, {Satoh}, {Ishitani Silva}, {Toda},
  {Tristram}, {Vandorou}, {Yama}, {Abrams}, {Agarwal}, {Rose}, \&
  {Terry}}]{lam2022}
{Lam}, C., {Lu}, J.~R., {Udalski}, A., {et~al.} 2022{\natexlab{a}}, \apjl, 933,
  L23

\bibitem[{{Lam} {et~al.}(2022{\natexlab{b}}){Lam}, {Lu}, {Udalski}, {Bond},
  {Bennett}, {Skowron}, {Mroz}, {Poleski}, {Sumi}, {Szymanski}, {Kozlowski},
  {Pietrukowicz}, {Soszynski}, {Ulaczyk}, {Wyrzykowski}, {Miyazaki}, {Suzuki},
  {Koshimoto}, {Rattenbury}, {Hosek}, {Abe}, {Barry}, {Bhattacharya}, {Fukui},
  {Fujii}, {Hirao}, {Itow}, {Kirikawa}, {Kondo}, {Matsubara}, {Matsumoto},
  {Muraki}, {Olmschenk}, {Ranc}, {Okamura}, {Satoh}, {Ishitani Silva}, {Toda},
  {Tristram}, {Vandorou}, {Yama}, {Abrams}, {Agarwal}, {Rose}, \&
  {Terry}}]{lam2022_supl}
{Lam}, C., {Lu}, J.~R., {Udalski}, A., {et~al.} 2022{\natexlab{b}}, \apjs, 260,
  55

\bibitem[{{Liebes}(1964)}]{liebes1964}
{Liebes}, S. 1964, Physical Review, 133, 835

\bibitem[{{Miyamoto} \& {Yoshii}(1995)}]{miyamoto1995}
{Miyamoto}, M., \& {Yoshii}, Y. 1995, \aj, 110, 1427

\bibitem[{{Nemiroff} \& {Wickramasinghe}(1994)}]{nemiroff1994}
{Nemiroff}, R.~J., \& {Wickramasinghe}, W.~A.~D.~T. 1994, \apjl, 424, L21

\bibitem[{{Raghavan} {et~al.}(2010){Raghavan}, {McAlister}, {Henry}, {Latham},
  {Marcy}, {Mason}, {Gies}, {White}, \& {ten Brummelaar}}]{raghavan2010}
{Raghavan}, D., {McAlister}, H.~A., {Henry}, T.~J., {et~al.} 2010, \apjs, 190,
  1

\bibitem[{{Sahu} {et~al.}(2022){Sahu}, {Anderson}, {Casertano}, {Bond},
  {Udalski}, {Dominik}, {Calamida}, {Bellini}, {Brown}, {Rejkuba}, {Bajaj},
  {Kains}, {Ferguson}, {Fryer}, {Yock}, {Mroz}, {Kozlowski}, {Pietrukowicz},
  {Poleski}, {Skowron}, {Soszynski}, {Szymanski}, {Ulaczyk}, {Wyrzykowski},
  {Barry}, {Bennett}, {Bond}, {Hirao}, {Ishitani Silva}, {Kondo}, {Koshimoto},
  {Ranc}, {Rattenbury}, {Sumi}, {Suzuki}, {Tristram}, {Vandorou}, {Beaulieu},
  {Marquette}, {Cole}, {Fouque}, {Hill}, {Dieters}, {Coutures},
  {Dominis-Prester}, {Bennett}, {Bachelet}, {Menzies}, {Alb-row}, {Pollard},
  {Gould}, {Yee}, {Allen}, {de Almeida}, {Christie}, {Drummond}, {Gal-Yam},
  {Gorbikov}, {Jablonski}, {Lee}, {Maoz}, {Manulis}, {McCormick}, {Natusch},
  {Pogge}, {Shvartzvald}, {Jorgensen}, {Alsubai}, {Andersen}, {Bozza}, {Calchi
  Novati}, {Burgdorf}, {Hinse}, {Hundertmark}, {Husser}, {Kerins},
  {Longa-Pena}, {Mancini}, {Penny}, {Rahvar}, {Ricci}, {Sajadian}, {Skottfelt},
  {Snodgrass}, {Southworth}, {Tregloan-Reed}, {Wambsganss}, {Wertz}, {Tsapras},
  {Street}, {Bramich}, {Horne}, \& {Steele}}]{sahu2022}
{Sahu}, K.~C., {Anderson}, J., {Casertano}, S., {et~al.} 2022, \apj, 933, 83

\bibitem[{{Sch{\"o}nrich} {et~al.}(2010){Sch{\"o}nrich}, {Binney}, \&
  {Dehnen}}]{schonrich2010}
{Sch{\"o}nrich}, R., {Binney}, J., \& {Dehnen}, W. 2010, \mnras, 403, 1829

\bibitem[{{Skowron} {et~al.}(2016){Skowron}, {Udalski}, {Koz{\l}owski},
  {Szyma{\'n}ski}, {Mr{\'o}z}, {Wyrzykowski}, {Poleski}, {Pietrukowicz},
  {Ulaczyk}, {Pawlak}, \& {Soszy{\'n}ski}}]{skowron2016}
{Skowron}, J., {Udalski}, A., {Koz{\l}owski}, S., {et~al.} 2016, \actaa, 66, 1

\bibitem[{{Tomaney} \& {Crotts}(1996)}]{tomaney1996}
{Tomaney}, A.~B., \& {Crotts}, A. P.~S. 1996, \aj, 112, 2872

\bibitem[{{Walker}(1995)}]{walker1995}
{Walker}, M.~A. 1995, \apj, 453, 37

\bibitem[{{Witt} \& {Mao}(1994)}]{witt1994}
{Witt}, H.~J., \& {Mao}, S. 1994, \apj, 430, 505

\bibitem[{{Wo{\'z}niak}(2000)}]{wozniak2000}
{Wo{\'z}niak}, P.~R. 2000, \actaa, 50, 421

\end{thebibliography}

\begin{table*}
\caption{Results of the Fit to the Photometric OGLE Data Only}
\centering
\begin{tabular}{lrr}
\hline \hline
Parameter               & Old Reductions & New Reductions                  \\
\hline
$t_0$ ($\mathrm{HJD}'-5760$)  & $3.3258 \pm 0.0007$ & $3.3286 \pm 0.0006$ \\
$\tE$ (d)               & $247.4 \pm 2.5$ & $239.4 \pm 2.4$           \\
$u_0$                   & $0.00254 \pm 0.00003$ & $0.00263 \pm 0.00003$     \\
$\piEN$                 & $0.100 \pm 0.007$ & $0.080 \pm 0.009$         \\
$\piEE$                 & $-0.060 \pm 0.003$ & $-0.052 \pm 0.004$        \\
\multicolumn{3}{l}{Derived Parameters:} \\
$t_{\rm eff}$ (d)       & $0.6295 \pm 0.0008$ & $0.6299 \pm 0.0007$       \\
$\piE$                  & $0.116 \pm 0.007$ & $0.095 \pm 0.009$         \\
$\varphi$ (deg)         & $353.7 \pm 2.3$ & $345.1 \pm 3.7$           \\
$I_{S}$ (mag)           & $19.731 \pm 0.012$ & $19.815 \pm 0.011$        \\
$(\Fb/\Fs)_{\rm OGLE}$  & $20.34 \pm 0.22$ & $20.47 \pm 0.22$          \\
$(\Fb/\Fs)_{\mathrm{HST},F814W}$ \citep{sahu2022} & $0.063 \pm 0.010$ & $0.028 \pm 0.011$ \\
$(\Fb/\Fs)_{\mathrm{HST},F606W}$ \citep{sahu2022} & $0.057 \pm 0.011$ & $0.022 \pm 0.011$ \\
$(\Fb/\Fs)_{\mathrm{HST},F814W}$ \citep{lam2022}  & $0.159 \pm 0.015$ & $0.110 \pm 0.014$ \\
$(\Fb/\Fs)_{\mathrm{HST},F606W}$ \citep{lam2022}  & $0.104 \pm 0.013$ & $0.059 \pm 0.013$ \\
$\chi^2/\mathrm{dof}$ & $12107.8/12430$ & $12577.2/12493$          \\
\hline
\end{tabular}
\label{tab:fit}
\end{table*}

\begin{table*}
\caption{Results of the Joint Fit to the Astrometric \textit{HST} and Photometric OGLE Data}
\centering
\begin{tabular}{lrrrrr}
\hline \hline
Parameter               & Sahu+           & Lam+ & Sahu+ w/ Syst.          & Lam+ w/ Syst.\\
\hline
$t_0$ ($\mathrm{HJD}'-5760$)  & $3.3286 \pm 0.0006$ & $3.3285 \pm 0.0006$& $3.3286 \pm 0.0006$       & $3.3286 \pm 0.0006$\\
$\tE$ (d)               & $239.7 \pm 2.4$           & $238.7 \pm 2.2$    & $239.7 \pm 2.4$           & $239.1 \pm 2.4$\\
$u_0$                   & $0.00263 \pm 0.00003$     & $0.00264 \pm 0.00003$ & $0.00263 \pm 0.00003$  & $0.00263 \pm 0.00003$\\
$\piEN$                 & $0.073 \pm 0.007$         & $0.091 \pm 0.007$  & $0.075 \pm 0.008$         & $0.083 \pm 0.007$ \\
$\piEE$                 & $-0.050 \pm 0.003$        & $-0.056 \pm 0.003$ & $-0.051 \pm 0.003$        & $-0.053 \pm 0.003$ \\
$\mu_{\alpha}$ (mas/yr, source) & $-2.281 \pm 0.022$& $-2.250 \pm 0.020$ & $-2.283 \pm 0.027$        & $-2.226 \pm 0.042$ \\
$\mu_{\delta}$ (mas/yr, source) & $-3.624 \pm 0.021$& $-3.567 \pm 0.020$ & $-3.612 \pm 0.037$        & $-3.571 \pm 0.025$ \\ 
$\theta_{\rm E}$ (mas)  & $5.68 \pm 0.40$           & $3.65 \pm 0.65$    & $5.71 \pm 0.61$           & $4.02 \pm 0.81$ \\
$\sigma_x$ (mas)        & 0 (fixed)                 & 0 (fixed)          & $0.088^{+0.088}_{-0.061}$ & $0.433 ^{+0.138}_{-0.104}$ \\
$\sigma_y$ (mas)        & 0 (fixed)                 & 0 (fixed)          & $0.189^{+0.092}_{-0.072}$ & $0.124 ^{+0.130}_{-0.088}$ \\
\multicolumn{3}{l}{Derived Parameters:} \\
$\varphi$ (deg)         & $342.3 \pm 3.0$           & $349.8 \pm 2.7$    & $343.1 \pm 3.3$           & $346.5 \pm 3.1$\\
$\mu_{\alpha}$ (mas/yr, lens) & $-4.48 \pm 0.39$    & $-3.07 \pm 0.24$   & $-4.41 \pm 0.46$          & $-3.41 \pm 0.38$ \\
$\mu_{\delta}$ (mas/yr, lens) & $3.29 \pm 0.50$     & $1.02 \pm 0.83$    & $3.36 \pm 0.76$           & $1.42 \pm 1.00$ \\ 
$M$ ($M_{\odot}$)       & $7.88 \pm 0.82$           & $4.19 \pm 0.77$    & $7.80 \pm 1.09$           & $5.03 \pm 1.10$ \\
$D_l$ (kpc)             & $1.49 \pm 0.12$           & $1.79 \pm 0.28$    & $1.47 \pm 0.15$           & $1.77 \pm 0.31$ \\
$V$ (km\,s$^{-1}$)              & $-11.8 \pm 4.3$           & $-10.7 \pm 5.5$    & $-10.9 \pm 4.9$           & $-12.5 \pm 6.0$ \\
$W$ (km\,s$^{-1}$)              & $38.1 \pm 2.5$            & $22.6 \pm 5.6$     & $37.9 \pm 3.5$            & $26.5 \pm 6.7$ \\
$\ln\mathcal{L}$ & $-5776.3$ & $-5799.9$ & $-5774.6$ & $-5789.4$ \\
$\ln\mathcal{L}_{\rm phot}$ & $-6289.3$ & $-6289.5$ & $-6289.2$ & $-6289.1$ \\
$\ln\mathcal{L}_{\mathrm{astr}, x}$ & 259.5 & 236.2 & 259.5 & 245.9 \\
$\ln\mathcal{L}_{\mathrm{astr}, y}$ & 253.5 & 253.4 & 255.0 & 253.7 \\
\hline
\end{tabular}
\label{tab:fit3}
\end{table*}

\newpage

\appendix
\section{Blending from a Source Companion}
\label{sec:blending}

Source companions can in principle be at any separation, and therefore a detailed treatment should systematically consider them all. For example, if the source companion were within the Einstein ring, it should participate in the photometric event at some level, while if it is outside the \textit{HST} PSF, then it will not participate in the astrometric event at all. In the present treatment, we will consider that the companion is sufficiently separated that its orbital motion can be ignored, but sufficiently close that it is within the PSF. We will also assume that the astrometric microlensing of the companion induces negligible effects. We emphasize that, while these are plausible assumptions for many cases, if there are grounds for relaxing them, then the uncertainties induced by source--companion blending will be more severe than what we outline below. Our conclusions also hold true for neighbor stars unrelated to the source, in which case there are more free parameters, for example the proper motion of the blend.

It is reasonable to neglect the parallax of the source/blend because the source is almost certainly in the bulge, and the astrometric reference frame is completely dominated by bulge stars.  The parallax of the source relative to this frame is therefore $|\pi_{\rm rel,S-frame}| \la 40\,\mu$as, i.e., far smaller than can give
rise to measurable effects in the current data set. Then, the centroid of the light moves according to the equation
\begin{equation}
\boldsymbol{\theta}(t) = \frac{A(t)(\thetaE\delta\boldsymbol{u}(t)+\boldsymbol{\theta}_{S,0}) + r\boldsymbol{\theta}_{B,0}}{A(t)+r}+\boldsymbol{\mu}_S(t-t_0),
\end{equation}
where $r=\Fb/\Fs$ is the blend-to-source flux ratio, $\boldsymbol{\theta}_{S,0}$ and $\boldsymbol{\theta}_{B,0}$ are the position of the source and the blend at $t_0$, $\boldsymbol{\mu}_S$ is the proper motion of the source, and $\delta\boldsymbol{u}(t)$ is the astrometric microlensing centroid shift
\begin{equation}
\delta\boldsymbol{u}(t) = -\frac{\boldsymbol{u}(t)}{u^2(t)+2}.
\end{equation}
If we regard $r$ as externally determined, then this qquation is linear in its seven free parameters, i.e., two additional free parameters relative to the case in which blending is ignored ($r=0$).

We work in the limit $r\ll 1$, which implies that the astrometric displacement induced by the blend is
\begin{equation}
\delta\boldsymbol{\theta}(t) \approx \frac{r\Delta\boldsymbol{\theta}}{A(t)+r}\approx \frac{\boldsymbol{\epsilon}}{A(t)}.
\end{equation}
In our case, the companion separation $\Delta\theta< 30\,$mas, and we will assume $r=0.028$ (from Table~\ref{tab:fit}), i.e., $\epsilon<0.9\,$mas.
If we add such a time series of displacements due to blending (but interpret them incorrectly, without a blending term), it will change the magnitude and direction $\muhelio$ compared to their values if the position of the blend were known and its effects were properly taken into account.  We will show below, in particular, that the amplitude of this effect is large enough to explain the discrepancy in Figure~\ref{fig:pm1}, and also large enough to substantially change the inferred value of $\thetaE$, but not enough to alter the conclusion that a black hole has been detected.

We begin by presenting a completely general approach, in which the only two assumptions are that $r\ll 1$, and that the magnification as a function of time, $A(t)$, is known from the photometric observations.  This will provide a general framework for other cases.  We then introduce several simplifications that are justified for the present case.

We consider a time series of (2D vector) astrometric measurements, $\boldsymbol{z}(t_k)$, with (2D) covariance matrices (in some definite, adopted, coordinate frame) $C^k_{mn}$ and define the 2D matrix series $B^k\equiv (C^k)^{-1}$.  If these measurements are displaced by the blend by $\delta\boldsymbol{z}(t_k) = \boldsymbol{\epsilon}/A(t_k)$ but modeled without a blend, then the resulting parameters will be displaced by (e.g., \citealt{gould2003})
\begin{equation}
\delta a_{i,m} = \sum_{j=1}^3 \sum_{n=1}^2 c_{(i,m),(j,n)}d_{(j,n)},
\label{eq:ag1}
\end{equation}
where
\begin{equation}
c = b^{-1};
\quad 
b_{(i,n),(j,m)} = \sum_k B^k_{mn} f_i(t_k)f_j(t_k);
\quad
d_{(i,m)} = \sum_k B^k_{mn} f_i(t_k)\delta z^m(t_k)
\label{eq:ag2}
\end{equation}
and
\begin{equation}
f_1(t) = 1;
\quad
f_2(t) = t;
\quad
f_3(t) = -\frac{u(t)}{[u(t)]^2 + 2}.
\label{eq:ag3}
\end{equation}
Here, $\delta \boldsymbol{a}_1$, $\delta \boldsymbol{a}_2$, and $\delta \boldsymbol{a}_3$ are the spurious position, source proper motion, and ``vector angular Einstein radius'' $\boldsymbol{\theta}_{\rm E} = \tE \muhelio$ offsets induced by the blend.

If, in the adopted coordinate frame, the $C^k$ are diagonal, i.e., the astrometric measurements in the adopted ``$x$'' and ``$y$'' directions are uncorrelated, then Equations~(\ref{eq:ag1}) and~(\ref{eq:ag2}) decouple into two independent sets of equations
\begin{equation}
\delta a_i^m = \sum_{j=1}^3 c^m_{ij}d_j^m;
\quad
(m=1,2)
\label{eq:ag4}
\end{equation}
where
\begin{equation}
c^m = (b^m)^{-1};
\quad 
b_{ij}^m = \sum_k \frac{f_i(t_k)f_j(t_k)}{[\sigma^m(t_k)]^2};
\quad
d_i^m = \sum_k \frac{f_i(t_k)\delta z^m(t_k)}{[\sigma^m(t_k)]^2}.
\label{eq:ag5}
\end{equation}
Note, in particular, that this decoupling does not depend on the errors being isotropic: they could be very different in the ``$x$'' and ``$y$'' directions, as long as they are uncorrelated.

Next, we illustrate this effect for the present case, by evaluating the parameter displacement for the actual astrometric time series, assuming for simplicity that the measurement errors are all equal to each other (but still potentially different in the two directions). Because the parallax is small, we simply evaluate $A(u)=(u^2+2)/(u\sqrt{u^2+4})$. We find
\begin{equation}
\delta \boldsymbol{a}_i = (-0.035,0.093/t_{\rm E}, -2.44)\boldsymbol{\epsilon},
\end{equation}
i.e., $\delta \boldsymbol{\theta}_{\rm E}  = \delta \boldsymbol{a}_3 = -2.44\boldsymbol{\epsilon}$.
That is, the systematic displacement is completely independent of the scale of the statistical errors, and in particular, is isotropic, even if the scale of the errors is different in the two directions. 

Thus, retaining the assumed blending fraction $r=0.028$, if $\Delta\theta= 10\,$mas and the source companion lies due east of the source, then the {\it HST}-based astrometry measurement would be displaced to the west (i.e., negative eastward direction) from its ``true value'' (given by the OGLE photometric solution), by $\Delta\boldsymbol{\theta}_{\rm E} = -2.44\times 0.028\times 10\,{\rm mas}= -0.7\,$mas, which would approximately account for the discrepancy seen in Figure~\ref{fig:pm1}. However, if, as is equally possible, the same companion lies 10 mas east and 28 mas north or south of the source, then it would still explain the discrepancy in direction, but it would also cause $\thetaE$ to be $2.44\times 0.028\times 28\,{\rm mas}= 2.0\,$mas smaller or bigger.

The prior probability that the source has a companion that is within the factor $\sim 5$ range of projected separations and has sufficient brightness to affect the $\thetaE$ measurement at these levels is about 10\%. However, if there is a source companion, it would be at some random orientation, so that the angle between the ``apparent'' lens-source proper motion and the microlensing parallax direction would be some random number. Hence, it is unlikely that they would agree to within a few degrees, so the chance that the source companion is this close in angle is of order 2\%.
The empirical evidence that such a companion has been detected must be weighed in this context.

With regard to corruption by source--companion blends, OGLE-2011-BLG-0462 is actually a very favorable case.  The source is relatively bright, leading to an excellent parallax determination from the OGLE light curve (and so yielding a precise and small measurement of the blended light). In addition, because the source is bright, typical companions are faint relative to the source, leading to small $r$.  In a more typical microlensing event, the blend fraction could be much larger and still evade decisive measurement or even detection.

Thus, low-level blending by source companions is a major, previously unrecognized, challenge to astrometric microlensing measurements of black hole masses.

\end{document}